# Nanoporous Metals: From Plasmonic Properties to Applications in Enhanced Spectroscopy and Photocatalysis

*Alemayehu Nana Koya[1], Xiangchao Zhu[2], Nareg Ohannesian[3], A. Ali Yanik[2], Alessandro Alabastri[4], Remo Proietti Zaccaria[1,5], Roman Krahne[1], Wei-Chuan Shih[2], and Denis Garoli[1*]*

[1] Istituto Italiano di Tecnologia, *via* Morego 30, I-16163, Genova, Italy.
[2] Department of Electrical and Computer Engineering, University of California, Santa Cruz, CA 95064, USA
[3] Department of Electrical and Computer Engineering, University of Houston, Houston TX 77204, USA
[4] Department of Electrical and Computer Engineering, Rice University, Houston, 77005 TX, USA
[5] Cixi Institute of Biomedical Engineering, Ningbo Institute of Materials Technology and Engineering, Chinese Academy of Sciences, Zhejiang 315201, China

*Corresponding author: Prof. Denis Garoli, denis.garoli@iit.it;

**Abstract:** The field of plasmonics is capable of enabling interesting applications in the different wavelength ranges, spanning from the ultraviolet up to the infrared. The choice of plasmonic material and how the material is nanostructured have significant implications for ultimate performance of any plasmonic device. Artificially designed nanoporous metals (NPMs) have interesting material properties including large specific surface area, distinctive optical properties, high electrical conductivity, and reduced stiffness, implying their potentials for many applications. This manuscript reviews the wide range of available nanoporous metals (such as Au, Ag, Cu, Al, Mg, and Pt), mainly focusing on their properties as plasmonic materials. While extensive reports on the use and characterization of NPMs exist, a detailed discussion on their connection with surface plasmons and enhanced spectroscopies as well as photocatalysis is missing. Here, we report on different metals investigated, from the most used nanoporous gold to mixed metal compounds, and discuss each of these plasmonic materials' suitability for a range of structural design and applications. Finally, we discuss the potentials and limitations of the traditional and alternative plasmonic materials for applications in enhanced spectroscopy and photocatalysis.

Plasmonic nanostructures can confine free space electromagnetic energy into nanosized regions[1] and transform it into different forms including confined and scattering fields, high energy -'hot'- electrons and holes, or heat and thermal radiation. Depending on the application, nanostructures are designed, in principle, to mainly express one of such energy transformations. To this end, nanoporous metals have been recently introduced.[2] They are artificially designed metamaterials made of solid metals with nanosized porosity, ultrahigh specific surface area, good electrical conductivity, high structural stability, and tunable optical properties. This set of characteristics underlines their importance for many applications such as electrochemical and optical sensing,[3] photo and chemical catalysis,[4] and advanced energy technology.[5] The NPMs are truly synthetic, their structural and optical properties can be tuned by controlling preparation conditions and preparation strategies.

One of the fascinating characteristics of the NPMs is their optical properties associated with excitation of surface plasmon resonance (SPR).[6] Resonant excitation of the surface plasmons in metallic nanostructures often gives rise to tight confinement and enhancement of electromagnetic fields in ultrasmall volumes beyond the diffraction limit. NPMs coming in various structures (such as nanoparticles, nanorods, nanofilms, etc) are characterized by widely tunable localized surface plasmon resonance (LSPR) [7,8] that can be modeled using various techniques such as finite element method (FEM) or effective medium approximation (EMA). Since the characteristic sizes of the ligaments and pores in NPMs are usually much smaller than the wavelength of the impinging light($\lambda$), significant LSPR can be excited in a broad range of the electromagnetic spectrum spanning from the ultraviolet (UV) to the near-infrared (NIR). Because the SPR of the nanoporous metals can be easily manipulated either by tailoring pore sizes, modulating the metals' dielectric properties, or varying their dielectric environment, the optical properties of NPMs suggest promising applications in advanced spectroscopy, from ultraviolet to near-infrared regimes. [9,10,11] In particular, electromagnetic hotspots associated to NPMs with their electromagnetic enhancement, tightly confinement, and ease operability have widely been exploited, for instance applied to surface-enhanced Raman spectroscopy (SERS) [12] and fluorescence enhancement. [13]

Most studies on the plasmonic properties of nanoporous metals and their implications for spectroscopic applications have been predominantly based on the traditional coinage metals, particularly nanoporous gold (NPG) [14,15,16] and nanoporous silver (NPAg) [17,18,19] To a lesser extent, nanoporous copper (NPC) structures with tunable pore sizes and hierarchical geometries have also been investigated as a SERS substrate for efficient detection and ultrasensitive sensing applications. [20,21] Even though addressing slightly different applications, all these materials exhibit plasmonic properties from the visible to near-infrared range. However, emerging applications require expansion of nanoplasmonics toward the ultraviolet range. In this respect, most recently, Al, Rh and Mg have emerged as alternative materials very capable of operating in the UV regime. [22,23,24] The tunable plasmonic properties of a material resulting from their combination makes them highly promising choice for commercial applications. As a result, there is a concrete growing research interest to exploit UV plasmonic materials for various applications including deep ultraviolet (DUV) SERS and UV metal-enhanced fluorescence.

This review covers nanoporous metals' optical properties and applications ranging from the well-studied noble metals to recently emerging plasmonic materials. Particular emphasis is given to the current trends in the modeling of optical properties, structural characterization, and advanced spectroscopic applications of nanoporous metals made of Au, Ag, Cu, Al, Mg, and Rh. In terms of scope, the review covers broad range of topics that include material properties, spectral ranges, and practical implications (see Schematic 1). Based on these premises, this article is organized as follows: its introductory section is followed by a technical section on the main preparation strategies for NPMs. Then the manuscript reports a critical review of recent progress and state-of-the-art regarding modeling NPMs and their optical properties. This is succeeded by comprehensive overview of recent developments on implications of NPMs for single-molecule SERS, near-infrared sensing, metal-enhanced fluorescence, extraordinary optical transmission and photocatalysis. Finally, the future directions of NPMs are discussed and concluding remarks are forwarded.

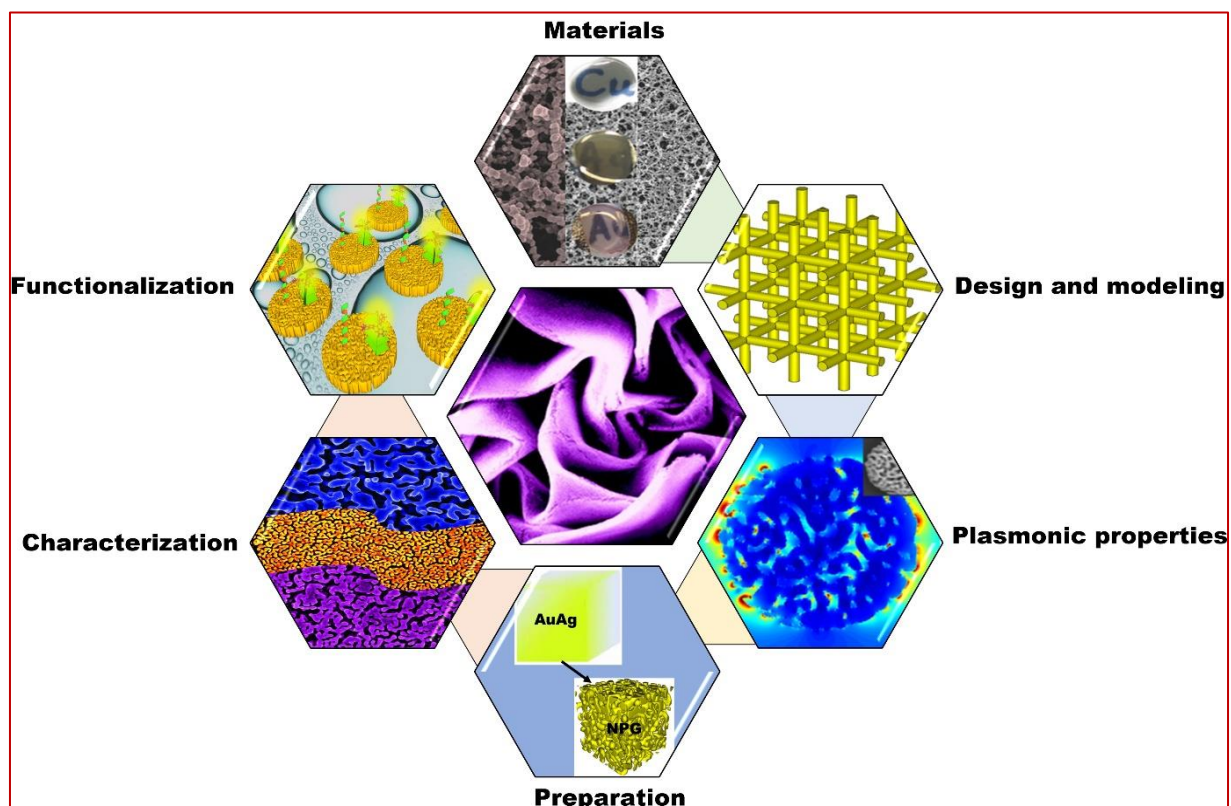

**Schematic 1**. Diversity of nanoporous metals (NPMs) including various material selection (Au, Ag, Cu, Al, Mg, Pt, Rh); design and modelling (finite element method(FEM), finite-difference time-domain (FDTD) method, effective medium approximation(EMA)); plasmonic properties (near-field properties and SPR spectra tuning); preparation strategies (dealloying, templating, galvanic replacement reaction, and physical vapour deposition); microcharacterization (Scanning electron microscopy(SEM), Tunneling electron microscopy(TEM), scanning transmission electron microscopy (STEM)); and functionalization of the NPMs (SERS, metal-enhanced fluorescence, and photocatalysis). Adapted with permission from ref. 25, Copyright 2011, American Chemical Society; ref. 26, Copyright 2018, WILEY-VCH; ref. 27, Copyright 2014, AIP Publishing LLC; ref. 28, Copyright 2020, Elsevier B.V; ref. 29, Copyright 2015, American Chemical Society; and ref. 30, Copyright 2016, American Chemical Society.

**Preparation Strategies and Plasmonic Properties of Nanoporous Metals**

The plasmonic properties of various metals including Au, Ag, Cu, Al, Mg, Pt, and Rh have been widely explored both experimentally and theoretically. The inner architecture of NPMs with random sizes and shapes can interact with the entire electromagnetic spectrum and result in excitation of SPRs. Thus, elucidating the fundamental plasmonic properties of NPMs and providing an accurate modeling of their responses is crucial for the efficient design of NPMs for applications in advanced spectroscopy. Moreover, since the optical and plasmonic properties of NPMs can be easily tuned by controlling the preparation conditions and strategies, it is crucial to overview recent developments in preparation strategies of nanoporous metals. In this section, current state-of-the-art in advanced preparation techniques, modeling methods, and optical properties of nanoporous metals in the spectrum from UV to NIR regimes are discussed.

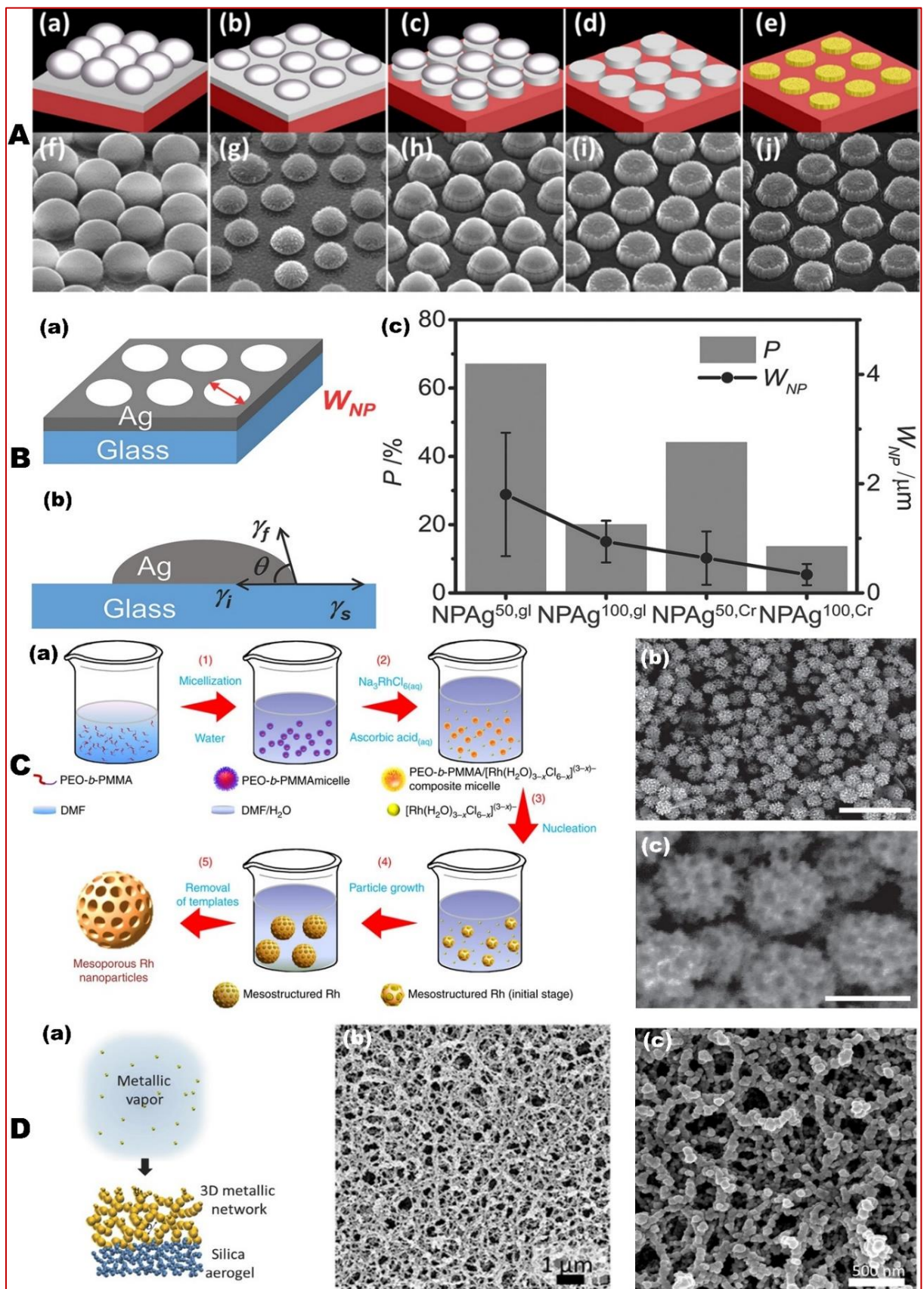

**Figure 1**. Preparation of nanoporous metals (NPMs). A) Fabrication steps to prepare nanoporous gold (NPG) disks using the combination of the lithographic pattering and atomic dealloying. (a) Formation of a monolayer of polystyrene (PS) beads on an alloy-coated substrate; (b) O$_2$ plasma shrinkage of the PS beads; (c) Ar sputter-etching to form isolated alloy disks; (d) Removal of PS beads; and (e) Formation of NPG disks by dealloying. (f–j) SEM Images taken at each process step. Adapted with permission from ref. 34, Copyright 2019, American

Chemical Society. B) Preparation of disordered NPAg thin films by thermally-assisted dewetting method. (a) Schematic representation of the NPAg film defined by a pore width, $W_{NP}$. (b) Schematic of Ag film dewetting mechanism that resulted in NPAg formation during the thermal annealing. According to the Young equation ($\gamma_s = \gamma_i + \gamma_f . cos\theta$), the metal remains as a continuous flat film only when surface energy of the bare substrate ($\gamma_s$) is larger or equal to the sum of the surface energy of the metal film ($\gamma_f$) and the substrate-metal interface energy ($\gamma_i$) at some contact angle θ between the silver surface and silver-substrate interface. (c) Porosities and pore widths of different NPAg sample types. Adapted with permission from ref. 36, Copyright 2015, WILEY-VCH. C) Preparation of mesoporous Rh nanostructures *via* chemical reduction on self-assemble polymeric poly(ethylene oxide)-b-poly(methyl methacrylate) (PEO-b-PMMA) micelle templates. Adapted with permission under a Creative Commons Attribution 4.0 International License from ref. 55, Copyright 2017, The Authors. D) Fabrication of nanoporous metallic networks based on the physical vapor deposition (PVD) strategy. (a) Illustration of the fabrication process of nanoporous metallic networks Vapored metallic atoms are directly self-organized into a 3D network of nanoscale features on top of the nanoporous silica aerogel substrate. The metallic vapor is produced by PVD, either by sputtering or evaporation. 3D SEM images of (b) gold and (c) silver networks. Adapted with permission from ref. 59, Copyright 2016, WILEY-VCH.

**Preparation of Nanoporous Metals**. Since nanoporous metals are truly synthetic, their plasmonic properties can be tuned by varying preparation conditions and strategies. Among several techniques for NPM preparation, templating, dealloying and colloidal chemistry are the most commonly utilized methods. Inspite of the fact that the template-based fabrication strategy gives much freedom to precisely control the size and microstructure of the final porous metal structures,[31] this technique is generally difficult and time consuming to implement. On the other hand, dealloying technique,[32] has been widely employed to fabricate various architectures including three-dimensional bicontinuous structures that are characterized by open nanopores, tunable pore sizes, structural properties and multifunctionalities.[2]

In particular, nanoporous metallic nanoparticles can be fabricated by combining the lithographic pattering technique and dealloying method. As shown in Figure 1A, nanoporous gold nanoparticles (for example, nanodisks [33,34]) have been fabricated by depositing gold and silver alloy onto a substrate made of, for example, silicon wafer or glass slide using an alloy target. A monolayer of polystyrene(PS) beads of different diameter is then can be formed on top of the alloy film. By selecting PS beads of various sizes, the diameter of the final NPG nanoparticles can be tuned. And then, to shrink the PS beads, a timed oxygen plasma treatment is often eployed, , which leads to separation of the PS beads from neighboring beads. To transfer the bead pattern into the alloy film, the sample is then sputter-etched in argon plasma. Once the pattern transfer is completed, the PS beads can be removedand the alloy disks are then dealloyed in concentrated nitric acid to produce an array of NPG nanoparticles (see Figure 1A(a)-(e)) . The SEM images shown in Figure 1A(f)-(j) show high-density NPG disk arrays on Si wafer before release. Such method is found to yield nanoporous metal particles with large surfe area, widely tunable surface plasmon resonance, and ultrahigh plasmonic hot spots that are essential for plasmon-enhanced applications [34, 35]. It should be noticed that NPG arrays can also be prepared using electron-beam lithography(EBL) over alternately deposited gold and silver layers through annealing process  The EBL technique can provide a large area array of nanoporous metallic particles with regular or random distribution and flexible interparticle separation [36]. Moreover, Yang and colleagues recently reported on development of a nanoporous silver structure with a tunable pore size and ligament using a silver halide electroreduction process .[35] Similarly, by employing a nonlithographic and thermally-assisted dewetting method, Shen and O'Carroll fabricated nanoporous silver films with different compositions to study their influence on the chain morphology and optical properties of conjugated polymers.[36] As illustrated in Figure 1B, NPAg can be formed by thermal annealing of a thin (50 nm – 100 nm) Ag film placed on (chromium (Cr) coated) glass (gl) substrate. The pores are formed owing to the difference in the surface energies between the thermally evaporated thin Ag film and the substrate. As shown in Figure 1B(c), the pore width ($W_{NP}$) and porosity (P) increases with decreasing the initial Ag film thickness from 100 nm to 50 nm. The addition of the Cr adhesion layer decreases both $W_{NP}$ and P, with more effect on the pore width than on the porosity.

In addition to the traditional coinage metals like Au and Ag, as a UV plasmonic material, Al has been extensively investigated by several authors.[22,37–49] Nanoparticles, nanoholes, nanopatterns and nanostructured films have been prepared to demonstrate enhanced spectroscopies in the UV spectral range. In almost the cases, the preparation of Al structures required several steps of process, from the chemical synthesis for the Al nanoparticles, to nanolithography for nanopatterned films. As for the other plasmonic metals, a nanoporous film can be an alternative approach to prepare plasmonic platforms with multiple LSPR hot-spots. To meet these growing demands, nanoporous aluminum nanostructures have been designed and prepared using various methods.[9,10,50–52] Recent papers have reported on fabrication of nanoporous aluminum substrates from an alloy of $Al_2Mg_3$ by means of a galvanic replacement reaction[10,51] intended for applications in enhanced UV Raman spectroscopy and fluorescence. Similarly, the same authors demonstrated the preparation of nanoporous Al−Mg (NPAM) alloy films by selective dissolution of Mg from a Mg-rich alloy $Al_xMg_{1-x}$.[9] The stoichiometry, porosity, and oxide contents in the NPAM can be tuned by modulating Al and Mg's ratio and the dealloying procedure. In addition, Okulov *et al.* also demonstrated the preparation of freestanding nanoporous Mg fabricated in a two-step process [55]. Intially, the Ti(Nb, Ta, V, Fe)$_{50}$Cu$_{50}$ alloys were dealloyed in liquid Mg in order to synthesize interpenetrating phase composites. In the second step, the Ti-rich phase was etched by selective dissolution in a 15 M aqueous solution of HF for several minutes in an ultrasonic bath, which is followed by cleaning in deionized water and alcohol. In another recent work, Liu *et al.* report on the preparation of nanoporous magnesium for hydrogen generation. In this case porous films can be prepared by means of physical vapor deposition starting from Mg powders with large granularity.[53]

Apart from the Al and Mg, rhodium (Rh) is another recently investigated material for UV plasmonics.[23,24,54–56] As for Al and Mg, some attempts in the preparation of nanoporous Rh structures have been reported. In particular, Jiang *et al.* illustrated the synthesis and characterization of mesoporous metallic rhodium nanoparticles (Figure 1C) [58]. They used a chemical reduction on self-assemble polymeric poly(ethylene oxide)-b-poly(methyl methacrylate) (PEO-b-PMMA) micelle templates (Figure 1C (a)). The PEO-b-PMMA micelles function as a soft template, while trisodium hexachlororhodate($Na_3RhCl_6$) served as the Rh precursor . Although in this case no plasmonic properties were investigated, the structural and morphological properties of the obtained nanoparticles (see Figure 1C (b & c)) suggest potential applications in plasmon-driven phenomena.

On the other hand, from the perspective of photocatalysis, in order to excite charge carriers that may be transferred to species in proximity and drive chemical transformation upon surface plasmon decay,[57,58] metallic nanostructures with a characteristic length less than 30 nm are needed. To meet this demand, Salmon group has reported on a simple and scalable method of fabricating pure nanoporous three-dimensional metallic networks.[26,59,60] This strategy is based on the physical vapor deposition (PVD) of vapored metallic atoms on a silica aerosol substrate that initiates self-assembly of the vapored metallic atoms into a nanoporous networks (Figure 2D). The resulting networks (made of various metals including Au, Ag, Cu, Al, Pt, Ti, and Fe) are found to be transparent, flexible and pure, implying their potentials for hot carriers generation and photocatalytic activity upon white-light illumination.[59]

**Modeling the Optical Responses of Nanoporous Metals**. To understand the fundamental optical and plasmonic properties and hence to predict the behaviors of the nanostructures, numerical modeling plays a key role. Among several simulation techniques,[61,62] finite element method (FEM)[63] and finite difference time domain (FDTD) method[33] have been employed to model the optical properties of NPMs. In FEM simulations, microscopic analyses from experimental samples (for example, SEM micrographs) can be imported as images on suitable platforms such as COMSOL Multiphysics® and used as permittivity maps by associating each points with a value between 0 and 1 according to the SEM brightness.[9,10,64] Similarly, although FDTD modeling is efficient for calculating far-field response (Figure 2A), the local field profile calculations are carried out by importing SEM images of porous metals.[33]

In the effective Medium Theory (EMT), which works for far-field optical parameters but cannot replicate near-filed and local effects,[11,27,65,66] all constituents are equally treated as fillers in a homogeneous medium that possesses the average properties of the composite. If the characteristic size of the constituents of the network is much smaller than the smallest optical wavelength in the respective material, the structure can be described as an effective medium [28]. That is, if there is a nanostructure comprised of two materials with respective volume fractions of $f_1$ and $f_2$ and if the electric field of the nanostructure is known, then we can easily deduce the effective permittivity $\varepsilon_{eff}$ of the medium as averaged electric displacement field D divided by the averaged electric field E, *i.e.,*[27]

$$\varepsilon_{eff} = \frac{\langle D \rangle}{\langle E \rangle} = \frac{f_1 \varepsilon_1 E_1 + f_2 \varepsilon_2 E_2}{f_1 E_1 + f_2 E_2} \tag{1}$$

where the sums of the averaged D-fields in the numerator and E-fields in the denominator, respectively, are weighed by their respective volume fraction, $f_1$ and $f_2$.

However, since the effective permittivity is not isotropic, eq 1 is not valid for all polarizations. Moreover, the given equation also assumes that electric fields in the constituting materials are constant, which is not generally true. Thus, the structure can be decomposed into smaller domains where electric fields are assumed to be approximately constant. This can be illustrated by developing a model system made of a cubic array of Au nanowires with dielectric voids in between. Such a model system can be divided into two separate sub-models, with the nanowires in parallel and perpendicular orientations. The effective permittivities of three such arrangements shown in Figure 2B (a) can be written as[27]

$$\varepsilon_{eff,\parallel} = \frac{f_{Au,\parallel}\varepsilon_{Au}E_{Au,\parallel} + f_d\varepsilon_d E_d}{f_{Au,\parallel}E_{Au,\parallel} + f_d E_d} = \frac{f_{Au,\parallel}\varepsilon_{Au} + f_d\varepsilon_d}{f_{Au,\parallel} + f_d} \tag{2}$$

$$\varepsilon_{eff,\perp} = \frac{f_{Au,\perp}\varepsilon_{Au}E_{Au,\perp} + f_d\varepsilon_d E_d}{f_{Au,\perp}E_{Au,\perp} + f_d E_d} = \frac{f_{Au,\perp}\varepsilon_{Au}\frac{2\varepsilon_d}{\varepsilon_{Au}+\varepsilon_d} + f_d\varepsilon_d}{f_{Au,\perp}\frac{2\varepsilon_d}{\varepsilon_{Au}+\varepsilon_d} + f_d} \tag{3}$$

$$\varepsilon_{eff,\#} = \frac{f_{Au,\perp}\varepsilon_{Au}E_{Au,\perp} + f_{Au,\parallel}\varepsilon_{Au}E_{Au,\parallel} + f_d\varepsilon_d E_d}{f_{Au,\perp}E_{Au,\perp} + f_{Au,\parallel}E_{Au,\parallel} + f_d E_d} \tag{4}$$

where the subscripts ∥ and ⊥ represent the media consisting of parallel and perpendicular wires, respectively, whereas # refers to the full cubic grid medium. $E_{Au,\parallel}$ and $E_{Au,\perp}$ are the fields in the respective sets of Au wires, $E_d$ is the field in the surrounding dielectric, $\varepsilon_{Au}$ and $\varepsilon_d$ are the wavelength-dependent permittivities of Au and dielectric and, finally, $f_{Au,\parallel}$, $f_{Au,\perp}$ and $f_d$, are the volume fractions occupied by the respective materials.[28] The effective permittivity given in eq 4 is a complete description of the full cubic grid medium, assuming that the structure is isotropic due to its cubic symmetry. It is obtained by averaging the fields over the parallel and orthogonal wires as well as the dielectric. By approximating NPM film by the effective medium model of a cubic grid of metal network, Jalas *et al.* investigated the spectral properties of nanoporous gold in the visible spectrum.[27] The transmission spectra of the effective medium for various permittivities of the embedding medium show two distinct peaks (Figure 2B (b)). The transmission deep can be further pronounced by increasing the gold content while keeping the dielectric permittivity in the voids constant. Similarly, by employing the Bruggeman effective medium theory, Ramesh *et al.* also modelled the optical reflectance of nanoporous gold film with isotropic pores in the near-infrared regime.[67]

However, the effective medium approximation is not valid for relatively larger pore sizes, since it only considers the effect of filling fraction by assuming the characteristic length is much smaller than the wavelength of interests. A recent report by Garoli *et al.* discussed a detailed investigation on the optical properties of nanoporous gold modeled the effective-medium approximation (EMA) .[12] The dielectric functions of two materials (vacuum and bulk gold) were combined in an analytical relation that defines the effective dielectric function. Different assumptions can be made regarding the local depolarization fields according to the different hypotheses that can be formulated on the inhomogeneous material structure. Concerning the previously mentioned works, in ref. [11] three different EMA models have been investigated: The Maxwell-Garnett EMA, [68] Landau−Lifshitz− Looyenga (LLL)[69] and Bruggeman EMA. [8] The Maxwell-Garnett typically works well for isolated particles of one strongly absorbing material dispersed in a weakly absorbing continuous matrix. In the LLL approach, the connectivity among the ligaments becomes the key parameter. In this case, the model works at the filling-factor limit. Both these two methods fail to model the optical response of NPM in the near-infrared. On the contrary, as previously reported, the Bruggeman formulation enables to model the main physical phenomenon in the porous metal. However, this method cannot quantitatively describe the electrodynamic response of the material because the complex material structure is not described by the filling factor alone[11]. Probably the most reliable method to model the optical response of NPMs is based on Drude-Lorentz model for complex permittivity $\varepsilon(\omega)$. This model has been used in several papers to obtain good fits to experimental reflectance and transmittance data from different nanoporous gold samples analyzed from the visible to the near-infrared spectral range.[11,70,71] The complex permittivity can be written as:

$$\varepsilon(\omega) = \varepsilon_\infty - \frac{A_D}{\omega^2 + i\omega\gamma_D} + \frac{A_L}{\pi}\frac{\gamma_L/2}{(\omega-\omega_L)^2 + \gamma_L^2} \qquad (5)$$

where $A_L$, $\omega_L$, and $\gamma_L$ are the strength, frequency and width of the Lorentz term, respectively, and $A_D$ and $\gamma_D$ are the intensity and width of the Drude term, respectively. This model enables to derive the dielectric constants of a generic NPM (Figure 2C). The dielectric constants of a metal are the fundamental parameters to evaluate its plasmonic properties.

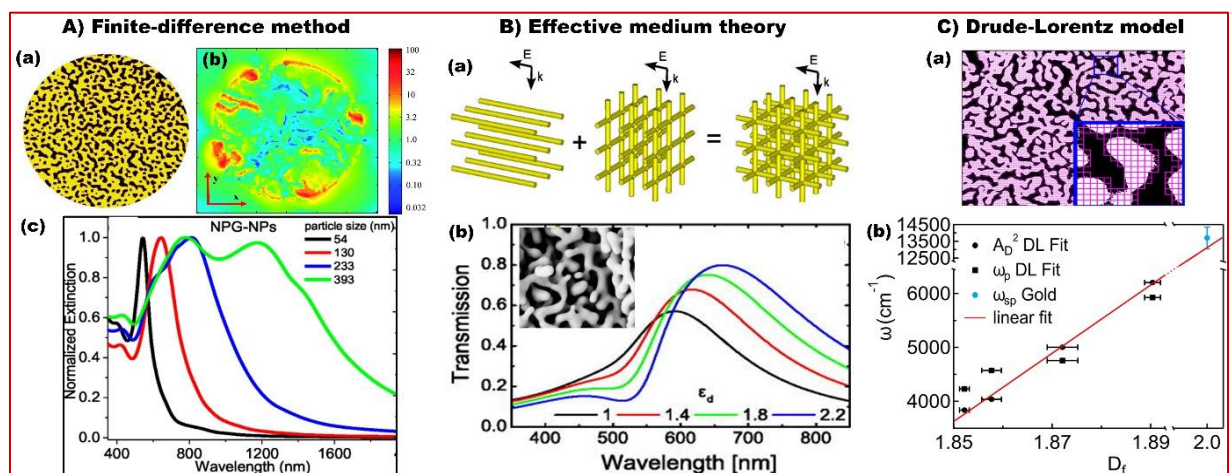

**Figure 2**. Modelling the optical responses of nanoporous metal particles, networks and films. A) Finite-difference time domain (FDTD) method-based modelling of nanoporous metal particles. (a) Schematic illustration of nanoporous gold (NPG) disk. (b) FDTD simulated electric field distribution of NPG disk with 500 nm diameter and 75 nm thickness. Adapted with permission from ref. 33, Copyright 2014, Nanoscale, the Royal Society of Chemistry. (c) FDTD calculated extinction spectra of nanoporous gold nanoparticles (NPG –NPs) with various particle sizes, 66% particle volume porosity, and 20 nm pore size. Adapted with permission from ref. 72, Copyright 2017, ACS Appl. Mater. Interfaces, American Chemical Society. B) Effective medium approximation (EMA) of nanoporous metal films by a cubic grid of gold wire model. (a) The model system consisting of a cubic grid of gold wires with two sub-models; the first sub-model considers only wires oriented parallel to the electric field, whereas the second sub-model considers the orthogonal wires. When linearly polarized wave illuminates a thin film of the model material under normal incidence, the wires oriented parallel to E-field vector contribute differently to the spectrum as compared to the orthogonal ones. (b) Calculated transmission through the effective medium of the cubic gold network for various permittivities of the embedding medium. The inset shows SEM of typical nanoporous gold film. Adapted with permission from ref. 27, Copyright 2014, Appl. Phys. Lett., AIP

Publishing LLC. C) Drude-Lorentz (DL) modelling of nanoporous gold film. (a) Fractal analysis of the SEM image of gold film. The fractal dimension ($D_f$), which can be directly computed from SEM images using the box-counting method that assigns a 0 or 1 value to each pixel in the SEM image, is the key morphological parameter to predict the plasmonic properties of NPG and simply tune them at will with the dealloying time. (b) Connection between the effective Drude model and the fractal analysis. Adapted with permission from ref. 11, Copyright 2018, ACS Photonics, American Chemical Society.

**Optical Properties of Nanoporous Metals**. The permittivity of a material is the fundamental quantity to evaluate its plasmonic properties. In addition, some key parameters such as (electromagnetic field) skin depth, propagation length and confinement and (resonance) quality factors are useful metrics to determine the material efficiency for plasmonic applications.[73] The plasmonic properties of NPMs are strictly related to the density of plasmonic hotspots that arise from surface plasmon resonances in the metals and their alloys. [74,75]

*Nanoporous Gold*. Nanoporous gold has been used as a model plasmonic metal to investigate the relationship between various parameters and porosities that are obtained by dealloying processes. In this respect, a plasmonic metal with skin depth up to hundreds of nm, propagation length up to 10 μm and a significant enhancement in the quality factor was recently introduced.[76] Furthermore, the relationship between the material electron temperature and its porosity has also been demonstrated. [76] As a result, NPG is the most extensively explored material owing to its chemical stability as well as optical, catalytic, and mechanical properties. [77] Among several porous structures, the plasmonic properties of NPG films have been widely investigated [7,8,78,79] intended for enhancements of SERS, [14,80] fluorescence emission [13,81] and electrochemical and optical sensing. [3] Extensive reviews on NPG and their application in plasmonic and other fields can be found in literature. [82,83,84,85]

Thin (∼100 nm) self-standing NPG membranes can exhibit both the propagating SPR excitation in the form of planar metal films and the localized SPR excitation in nano-featured metal architectures. [70,86] The propagating SPR spectrum shows a strong wavelength dependence leading to sharper SPR deep and smaller deep angle for longer excitation wavelength, which can be ascribed to efficient propagation of the SPR at the interfaces of the NPG film and dielectric. On the other hand, typical localized SPR spectra of NPG usually have a wide plateau with two characteristic peaks, which is distinctive and has not been observed in the LSPR spectra of other Au nanostructures. As displayed in Figure 3A(a)&(b), the two peaks show different plasmonic responses as the characteristic length (for example, pore size) of the NPG film is tuned from 10 nm to 50 nm. The peak position of the localized SPR at the shorter wavelength ($\lambda_1$) is not sensitive to the change of the nanopore sizes (Figure 3A(c)), suggesting that this short wavelength LSPR band may originate from the resonant absorption of the gold film. On the other hand, the LSPR band at the long wavelength ($\lambda_2$) results in a significant redshift as the pore size increases, with characteristic spectral feature compared to other nanostructured gold (for example, compare with the red dotted line of gold nanorod shown in Figure 3A(c)). This distinctive plasmonic behavior of the NPG can be attributed to the effect of radiation damping with the oscillation length of conduction electrons. However, Jalas and co-workers argued that, the two distinct peaks observed in the transmission spectra of NPG films could not be attributed to two separate localized surface plasmon resonances. [27] They claim that, the peculiar spectral feature of NPG films can be understood as that of diluted gold with a spectrally narrow dip in transmission due to the averaged electric field approaching zero, suggesting that the transmission characteristics are rather featured by a dip in one broad transmission curve than by two distinct peaks.

NPG typically containing a network of voids that occupy over 70% of the film volume behaves like a uniform Drude metal at NIR frequencies, but it is plasmonic at long wavelengths.[87] Although NPG is well known to be plasmonic in the NIR spectral range, the majority of the works investigating its plasmonic properties focused on the spectral range between 500 and 900 nm where the metal can generate LSPR, [7,8,71,88] which has been extensively exploited for SERS and biosensing, as it will be discussed later. On the contrary, detailed analyses on the plasmonic and optical properties in the infrared spectral region have been reported only in few cases. [11,87] In particular, the dependency of the plasma frequency of the NPG material with respect to the nanoporosity has been recently reported. The plasma frequency is a key parameter in the design of NPG-based metamaterials where the material switches from a metallic behavior to a high absorber film. More interestingly, the tunability in the plasma frequency can be applied to realize highly sensitive biosensors.[89]

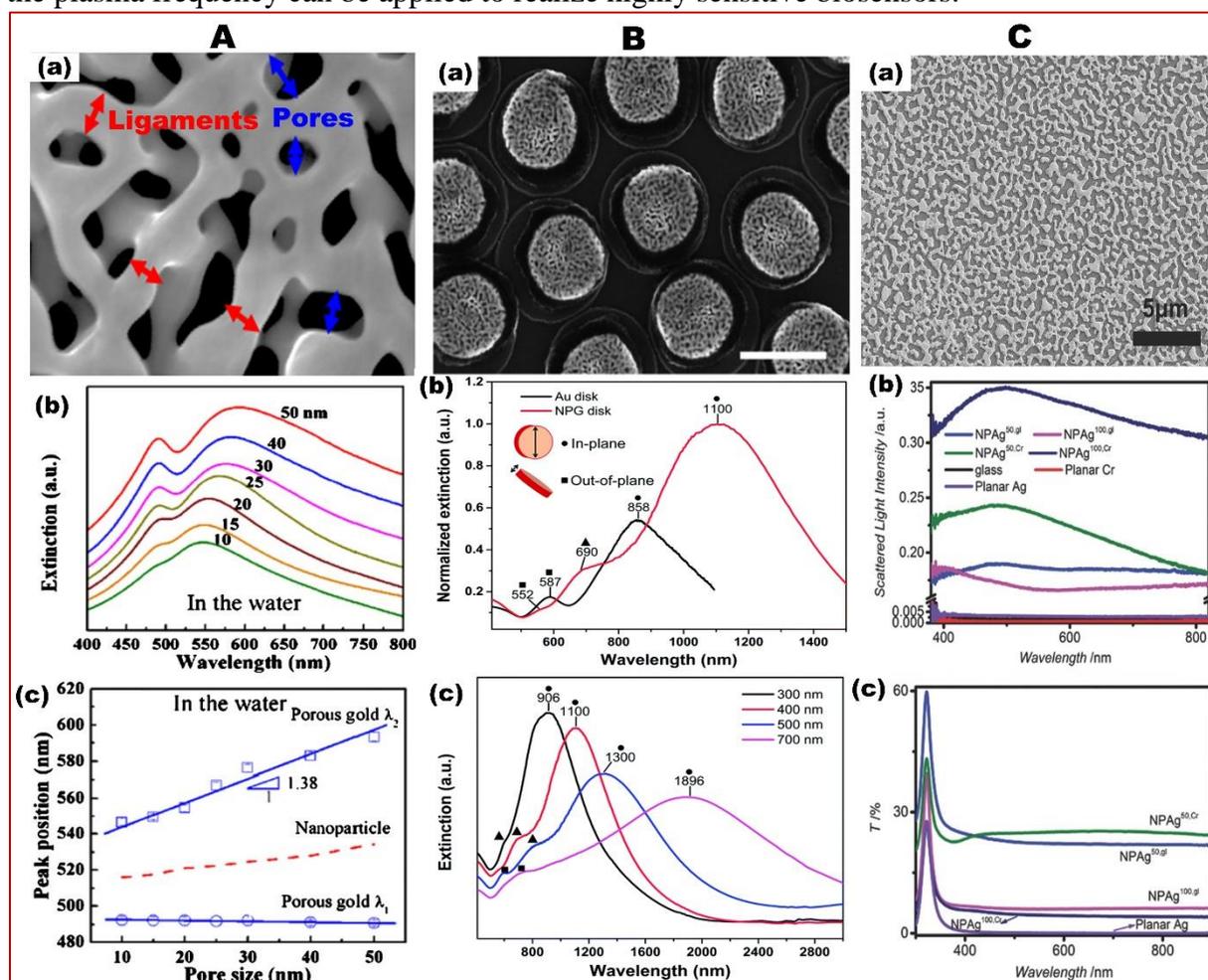

**Figure 3**. Tunable plasmonic properties of nanoporous gold (NPG) and nanoporous silver (NPAg) structures. A) Tuning localized surface plasmon resonance (LSPR) of thin NPG film. (a) Illustration of the average sizes of ligaments and pores in typical NPG film. Adapted with permission from ref. 8, Copyright 2014, AIP Publishing LLC. (b) Extinction spectra and (c) corresponding two resonant peak positions ($\lambda_1$ and $\lambda_2$) of NPG film with the pore sizes of 10 nm – 50 nm in water. The dashed red line in (c) represents the size dependence resonant band of gold nanoparticles. Reproduced with permission from ref. 7, Copyright 2011, AIP Publishing LLC. B) Plasmonic properties of NPG disks with tunable plasmon resonances. (a) SEM image of high-density NPG disks with diameters of 400 nm fabricated on Si substrate (with the scale bar of 500 nm). (b) Extinction spectra of 400 nm diameter and 75 nm thick Au disks and NPG disks on glass substrates measured in air. The inset shows the in-plane and out-of-plane resonance modes. (c) Size-dependent extinction spectra of NPG disks with different diameters (300, 400, 500, and 700 nm) consisted of high-density NPG disk monolayers on glass substrates in air. Reproduced with permission from ref. 33, Copyright 2014, the Royal Society of Chemistry. C) Optical properties of NPAg films. (a) Representative SEM image of the NPAg formed by annealing a 50-nm-thick Ag film on chromium-coated glass substrate. (b) Scattering and (c) transmittance spectra of different NPAg films. Adapted with permission from ref. 36, Copyright 2015, WILEY-VCH.

However, NPG exhibits weak plasmonic extinction and little tunablity of the surface plasmon resonance, owing to the fact that the pore size is much smaller than the wavelength of the light. As a result, recently, patterned NPG nanoparticles have emerged as better substrates for high-performance SERS applications, [90,33] as well as other forms of plasmon-enhanced spectroscopy such as near-infrared absorption and fluorescence. [91,92] NPG nanoparticle arrays are also highly effective in photothermal conversion for microfabrication, microfluidic manipulation, and pathogen inactivation.[93,94,95,96,97] As shown in Figure 2A, disk-shaped NPG nanoparticles can be fabricated directly on different substrates, making them highly reproducible nanostructures with exceptionally good robustness for integration with microfluidic nanodevices and surface functionalization.An abundance of plasmonic hot spots where the electric field is highly enhanced and easily accessible can be formed in the vicinity of the nanopores. More importantly, NPG nanoparticles have a much larger surface area compared to other existing plasmonic entities that is actually not due to the aggregation of high-density individual smaller nanoparticles (see Figure 2A(b) & 3B(a)). Instead, NPG nanoparticles derive the nanoporosity from atomic dealloying or selective dissolution of the less noble metal out of a binary alloy such as gold-silver alloy [98,99,100] as they are fabricated by the combination of top-down patterning and bottom-up atomic dealloying. The patterning step provides additional knobs to tailor the LSPR properties by defining a nanoparticle shapeand additional LSPR design parameters can be obtained by further shape engineering. By taking advantage of the abundant hot-spots and tunable plasmon resonance in NPG nanoparticles, a number of label-free sensors have been developed, including single-molecule DNA hybridization monitoring, [101] molecular sensing and imaging, [102] integrated microfluidic SERS sensor, [103], and sensing of malachite green and telomerase[104,30].

Unlike the NPG thin films that exhibit weak light-matter interaction and limited tunability, patterned high-density disk-shaped NPG nanoparticles acquire a prominent SPR. As shown in the extinction spectra in Figure 3B(b), the three resonance modes of the NPG disk can be assigned as NPG LSPR (▲), out-of-plane resonance (■), and in-plane resonance (●) [34]. The NPG LSPR mode originates from the nanoporous structures, whereas the in-plane and out-of-plane modes are associated with the external nanoparticle shape. Among these peaks, the in-plane resonance clearly dominates and only exists in NPG disks but not in semi-infinite NPG thin films (Compare Figure 3A(b) & 3B(b)). The size-dependent plasmonic shifts of these peaks can be observed when the disk diameter is increased from 300 to 700 nm, leading to notable tuning of the SPR from ~ 900 nm 1850 nm (Figure 3B(c)) . Finally, it is worth noticing that the LSPR peaks of NPG disk array exhibita significant red shift compared to that of individual NPG disk, which can be attributed to far-field radiative coupling from neighboring NPG disks. [105]

*Nanoporous Silver*. Similarly to NPG, the potentials of nanoporous silver have been extensively exploited for various plasmon-enhanced applications including SERS single-molecule detection, [18,19,106] fluorescence amplification, [107] catalysis, [108] and next-generation energy storage devices. [35] In particular, NPAg frameworks with characteristics of, for example, large specific surface area, electric conductivity, and porosity, are desirable for metal oxide-based pseudocapacitors, implying their potentials as a promising candidates for the next generation energy storage devices. On the other hand, disordered NPAg thin films (Figure 3C(a)) based on conjugated polymers (see Figure 2B & 3C) demonstrate potentials to influence chain morphology and optical properties of the conjugated polymers in optoelectronic devices. [36] The optical properties of the NPAg films displayed in Figure 3C(b) show strong back-scattering for all NPAg morphologies with small pore sizes. To be specific, the scattering intensity of the nanoporous film is the most intense for the NPAg films prepared on chromium coated glass whereas the bare glass substrate-based NPAg films exhibit weaker scattering intensities, which can be ascribed to the effects of the pore width and porosity. These broadband scattering spectra with their peaks in the 400 nm – 550 nm wavelength range are attributed to LSPRs of the randomly-organized nanostructures and out-coupled surface plasmon polaritons (SPPs). Similarly, the transmittance spectra of various samples displayed in Figure 3C(c) show resonance peaks around $\lambda \approx 320$ nm for all of the Ag-containing samples due to the transparency of Ag near its plasma frequency. [36]

*Other Metals*. Gold and Silver and their nanoporous configurations are the two most investigated plasmonic metals. However, several other metals have been explored as the primary material for nanoporous structure fabrications. Depending on the fabrication method, copper, aluminum, nickel, titanium, iron, palladium and platinum have been reported as NPMs.[26,109] In particular, copper has been extensively investigated as nanoporous metal and, compared to gold and silver, it is inexpensive and naturally abundant. Recently, several papers reported on the plasmonic properties and applications of copper structures for various purposes .[20,21,110,111] Regarding nanoporous Cu, the investigations have been mainly focused on applications such as electrochemistry and catalysis.[112–117] On the contrary, the plasmonic properties of Cu and Cu nanoparticles have been reported by several authors and can be considered as the basis to explore the plasmonic behavior of nanoporous Cu film.[118] Song and co-workers employed a one-step dealloying method to fabricate free-standing hierarchical nanoporous copper (HNPC) membranes from $Mg_{72}Cu_{28}$ alloy as precursor. [21] The hierarchical architecture is composed of large porosity channels (sized 50–100 nm) with a number of smaller pores (less than 20 nm in size) on channel walls (see Figure 4A(a)), which offers combined advantages of highly accessible surface and high density of plasmonic hotspots . The elemental mapping in STEM images shown in Figure 4A (b-d) identifies that the bright regions correspond to Cu-rich $CuMg_2$ whereas the dark regions correspond to Cu-absent Mg phase. The presence of secondary pores and primary ligaments in such hierarchical nanoporous structures gives rise to a cooperative enhancement of the surface plasmon resonance upon photon excitations. The FDTD simulation results shown in Figure 4e unveil that the local field enhancement is highly concentrated in the boundaries, tips, and inner pore regions, and points with the strongest electromagnetic field are mostly located at the curvature of the primary ligaments. The potential of nanoporous Cu as a plasmonic platform has been mainly investigated with regard to SERS application and the dependency of the SERS enhancement by the pore size and material preparation has been reported.

Most recently, aluminum (Al) has emerged as a viable alternative to the traditional plasmonic metals (such as Au and Ag) due to its distinctively favorable dielectric properties. [119,22,120,121] Since its large plasma frequency leads to a negative permittivity (real part) down to the wavelength of 100 nm, [122,123] aluminum has been a promising plasmonic material for the ultraviolet regime (Figure 4B (a, b)). It also exhibits strong local field enhancement owing to high electron density These interesting plasmonic properties combined with its natural abundance, low cost, amenability to manufacturing processes and compatibility with optoelectronic devices makes aluminum a highly promising material for commercial applications [23]. As a result, there is a growing research interest to exploit Al as plasmonic material for various applications including UV nanoantenna, DUV surface-enhanced Raman spectroscopy, and UV metal-enhanced fluorescence.

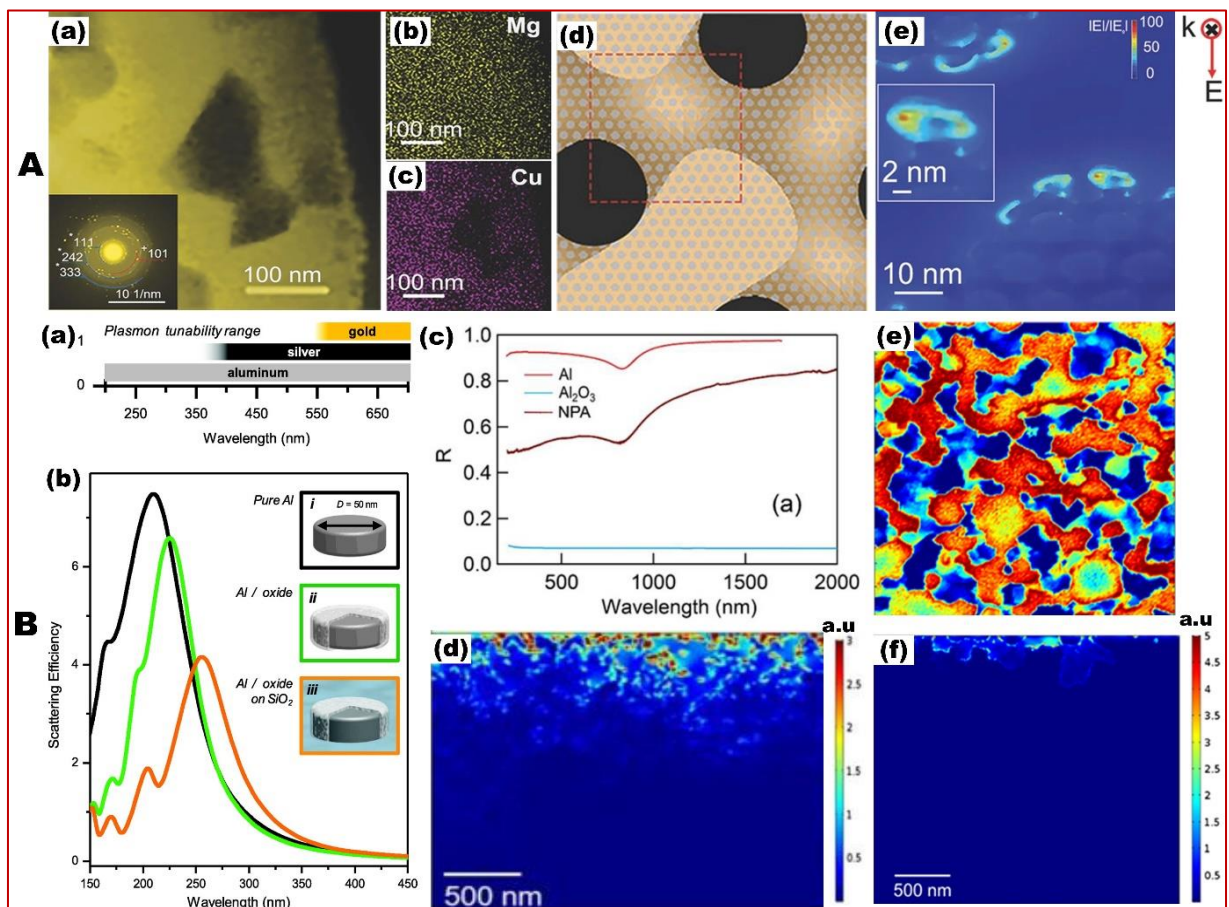

**Figure 4**. Optical and plasmonic properties of nanoporous copper (NPC) and nanoporous aluminum (NPA). A) Structural characterization and electric field distributions of hierarchical nanoporous copper (HNPC) structures. (a) Scanning transmission electron microscopy (STEM) image of $Mg_{72}Cu_{28}$ alloy ribbon, where the selected-area electron diffraction (SAED) of which shown in the inset demonstrates that (111), (242), and (333) correspond to $CuMg_2$ phase whereas (101) to Mg phase. (b, c) Element mappings elucidates that the dark phase in (a) is Mg while the bright phase is $CuMg_2$. (d) Schematic of the HNPC used to simulate electric field distributions. (e) Typical electric field distribution ($|E|/|E_0|$) of the HNPC on the top surface of the ligament with pore size of 10 nm. Reproduced with permission from ref. 21, Copyright 2018, WILEY-VCH. B) Optical and plasmonic properties of pure aluminum, aluminum oxides, aluminum alloys and nanoporous aluminum structures. (a) Plasmon resonance tuning ranges of the most common plasmonic materials, Au and Ag, compared with Al. (b) Calculated spectra for Al nanodisk (35 nm thick and 50 nm diameter) of (i) a pure Al, isolated Al nanodisk (black line); (ii) an isolated Al nanodisk with a 3 nm surface oxide (green); and (iii) the same Al nanodisk on an infinite $SiO_2$ substrate (orange). Reproduced with permission from ref. 22, Copyright 2013, American Chemical Society. (c) Reflectance spectra of pure Al, aluminum oxide, and nanoporous Al films. (d) Numerically-computed local field enhancement of NPA film calculated at an excitation wavelength of 260 nm. Reproduced with permission under a Creative Commons Attribution (CC BY) License from ref. 10, Copyright 2020, Nanomaterials. (e) Imported map of the horizontal cross section of the as-prepared nanoporous aluminum-magnesium alloy (NPAM). (f) Electromagnetic calculations of field confinement (a.u.) of the NPAM film calculated at an excitation wavelength of 260 nm. Reproduced with permission from ref. 9, Copyright 2019, American Chemical Society.

Garoli *et al.* have investigated the optical and plasmonic properties of nanoporous aluminum (NPA).[10] To evaluate the optical performances, they measured the reflectance from Al, $Al_2O_3$, and NPA samples in the spectral range spanning from 200 nm to 2000 nm with presence of absorption band at 800 nm (see Figure 4B (c)), which can be ascribed to the Al interband transition. Moreover, to explore the plasmonic properties of the NPA in the UV regime, they also computed the electric field enhancement and its spatial distribution by means of a 2D electromagnetic simulation (Figure 4B (d)). The same authors experimentally and numerically demonstrated that nanoporous Al-Mg (NPAM) alloy films show enhanced performances in the UV range.[9] They utilized experimental cross-sectional scanning electron microscopy (SEM) images of the fabricated NPAM films as the input geometry for electromagnetic calculations. This enables to highlight the effectiveness of NPAM in terms of electric field penetration and enhancement within the material pores (Figure 4B(e)). The electromagnetic local field distributions calculated by the finite element method show that the nanoporous Al-Mg alloy films have local field enhancement in the UV spectral range (Figure 4B (f)). It was also found that, compared to equivalent NPM films, the nanoporous Al−Mg alloy in the UV range demonstrated superior SERS and Metal Enhanced Fluorescence performances. The use of magnesium combined with aluminum is justified by similar behavior in the UV spectral range. Mg is now investigated by several authors as potential plasmonic material also with tunable properties thanks to its reversible oxidation.[24,124–126] Recently examples of nanoporous Mg films have been reported. Unfortunately, no specific investigation on its plasmonic properties has been reported so far.

**Nanoporous Metals for Enhanced Spectroscopy**
To meet the growing demand of sensing platforms that operate in UV, to the visible and up to the near-infrared wavelengths, NPMs have emerged as ideal sensing substrates for rapid, quantitative readout of the smallest amount of analytes with excellent specificity and high sensitivity. With such sensing platforms, it has been possible to detect single molecules with high accuracy and optimal performance. Thus, here, we illustrate the recent developments on nanoporous metal-enhanced biosensing and detection in broad spectral regimes.

**Surface-enhanced Raman Spectroscopy**. Plasmonic nanostructures are capable of confining electromagnetic fields into deep-subwavelength volumes with ultrahigh local field enhancement.[127] These hot spots have been widely exploited for enhancing the Raman signals of small molecules.[128] Generally, the electromagnetic enhancement of Raman signals of probe molecules depends on the intensity of total electric field at the molecule position. The EM field-induced SERS enhancement factor (SERS EF) is proportional to the fourth power of the near-field intensity enhancement |E|(ratio of the localized EM field at the location of the analyte molecule to the incident excitation field),[129] *i.e.*

$$SERS\ EF = |E|^4 = |\frac{E_{loc}}{E_0}|^4 \tag{6}$$

where $E_{loc}$ and $E_0$ are the electric field amplitudes at the structure surface and that of the incident light, respectively. However, for many SERS applications and experiments, it is important to consider the detailed distribution of the molecules on the SERS substrate. Therefore, it is necessary to define SERS substrate enhancement factor (SSEF), which can be used to compare the average SERS enhancements across different substrates.[130] The most widely used definition for the average SERS EF is:

$$SSEF = (I_{SERS}/N_{SERS})/(I_{RS}/N_{RS}) \tag{7}$$

where $I_{SERS}$ and $I_{RS}$ are the SERS and normal Raman intensities, respectively, $N_{SERS}$ is the number of probe molecules contributing to the SERS signal, and $N_{RS}$ is the number of the probe molecules contributing to the bulk Raman signal. For single-molecule (SM) SERS enhancement, the above expression can be simplified as $SMEF=I_{SERS}/I_{RS}$. As implied in eq (6), the electromagnetic SERS enhancement strongly depends on the hot spot intensity of SERS substrate, which can be further optimized by engineering the surface of the substrates. In particular, surface modification of porous metal substrates is crucial for triggering substantial local electromagnetic field enhancement around the roughened surfaces.[131] The NPM morphology and feature size can be tuned with material and process parameters such as, for example, alloy composition, or dealloying temperature and time. Almost all the investigated NPMs have been tested as SERS substrate. Nanoporous copper, silver, aluminum and others have been reported as potential platforms for SERS, with different enhancements related to the preparation procedures.[3,9,18–20,26,111,132–135] With respect to other metals, NPG has been much more investigated as platform for enhanced spectroscopy, in particular for SERS.

One of the well-known mechanisms of modifying the surfaces of NPMs films is by introducing 3D quasi-periodic wrinkles. This can be done by thermal contraction of pre-strained polymer substrates (PS) [136,137,138,139] which leads to formation of surface patterns (see Figure 5A(a)). Particularly, Zhang *et al.* demonstrated dramatic amplification of Raman intensity using wrinkled NPG films. These films contain a large number of Raman-active nanogaps produced by deformation and fracture of nanowire-like gold ligaments, which yield ultrahigh SERS for molecule detection. [25] Similarly, Liu and co-workers [140] demonstrated large-scale and chemically stable SERS substrate made from wrinkled nanoporous $Au_{79}Ag_{21}$ films that contain a high number of electromagnetic hot spots with a local SERS enhancement larger than $10^9$. Before the wrinkling treatment the film is flat (Figure 5A(b)), after the annealing quasi-periodic wrinkles are formed and distribute uniformly across the entire film also producing rose-petal-shape nanostructures (Figure 5A(c)). The Raman spectra of crystal violet (CV) solutions displayed in Figure 5A(d) & (e) reveal that wrinkling leads to significant improvement in the SERS signals compared to that of the flat NPG film. The SERS enhancements show strong pore-size dependence, where the wrinkled NPG (w-NPG) with the nanopore size of ~26 nm exhibits the highest SERS enhancement, about ~ 60-fold higher than that of the as-prepared NPG.[25] The excellent SERS performance of the wrinkled NPG film can be attributed to its heterogeneous nanostructures containing nano-pores, nano-tips, and nanogaps, which give the substrate a broad spectrum of plasmon frequencies for a wide range of molecule detection at a single-molecule level.

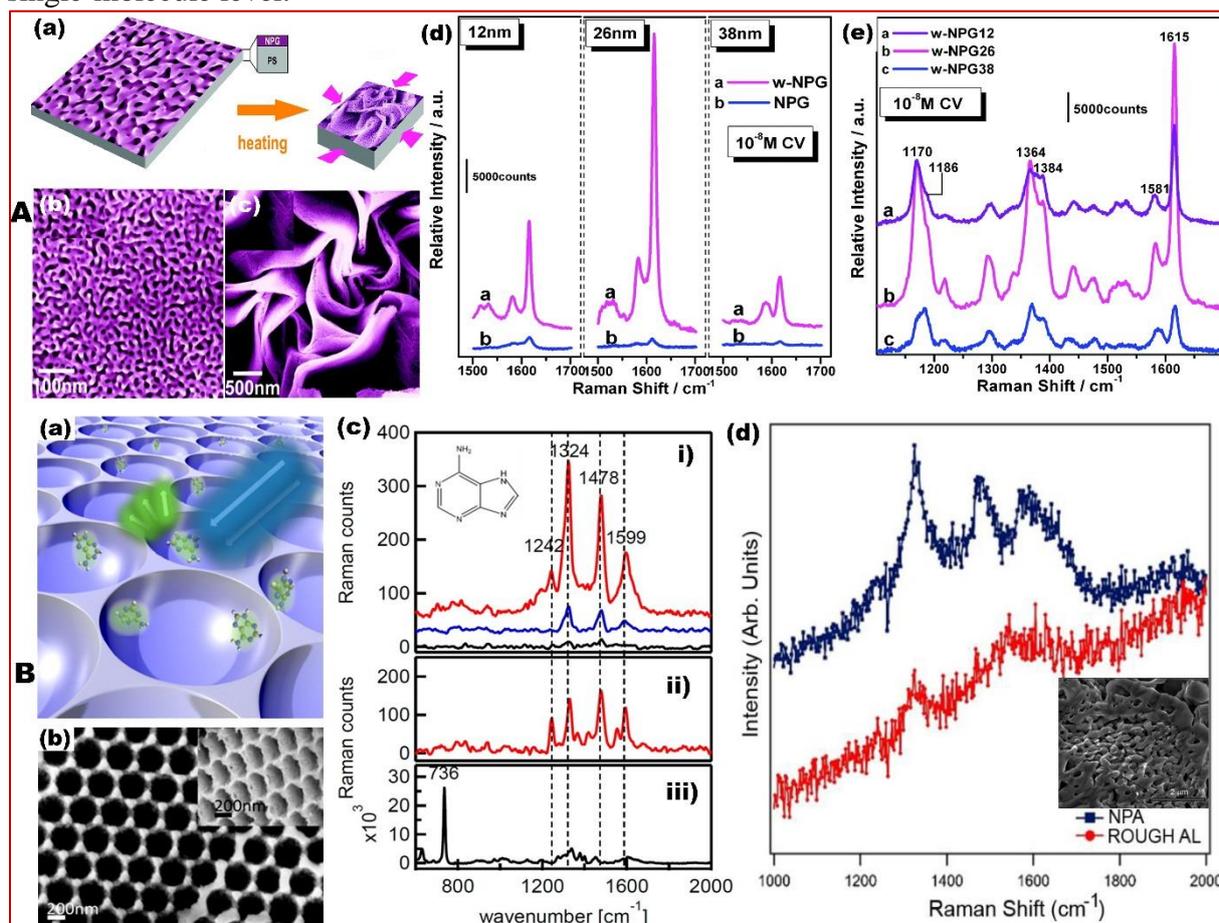

**Figure 5**. Nanoporous metal substrates for surface-enhanced Raman spectroscopy (SERS). A) Wrinkled nanoporous gold film SERS substrates. (a) Schematic diagram of the preparation of the wrinkled nanoporous gold film by thermal contraction of polymer substrate (PS). (b) Microstructure of as-prepared NPG and (c) zoom-in SEM micrograph of wrinkled NPG films with nanopore sizes of 12 nm. (d) Comparison of CV SERS spectra based on the wrinkled NPG (w-NPG) and as-prepared NPGs with different pore sizes. (e) SERS spectra from the wrinkled NPG with different pore sizes of 12 nm (w-NPG12), 26 nm (w-NPG26), and 38 nm (w-NPG38). The excitation wavelength is 632.8 nm. Reproduced with permission from ref. 25, Copyright 2011, American Chemical

Society. B) Aluminum for deep ultraviolet surface-enhanced Raman spectroscopy (DUV-SERS). (a) Schematic representation of nanovoid aluminum films. (b) SEM image of top view and 45° angled-view (inset) of the Al nanovoids. (c) (i) UV-SERS spectrum of a 1 mM adenine solution on a 200 nm void structured aluminum surface (red) compared to the UV-SERS spectrum on an evaporated aluminum surface (blue) and the resonant Raman spectrum of adenine solution without a plasmonic surface (black). The inset shows the structure formula of adenine. (ii) UV resonant Raman spectrum of bulk adenine in powder form. (iii) NIR SERS spectrum (excitation 785 nm) of adenine solution on Klarite. Reproduced with permission from ref. 153, Copyright 2013, American Chemical Society. (d) UV Raman spectra of salmon sperm DNA deposited on rough Al substrate (red curve) and nanoporous Al (NPA) substrate (blue curve). The inset show SEM of typical NPA film. Reproduced with permission under a Creative Commons Attribution (CC BY) License from ref. 10, Copyright 2020, Nanomaterials.

These demonstrations imply that the method the wrinkled NPG films is prepared is the most effective mechanism to achieve giant SERS enhancement. Anyway, many other approaches have been reported, from the optimization of the nanopore size,[15,14] to the coupling of NPG film with other metallic nanoparticles,[80,141] to the nanopatterning by means of imprinting[16,142] or sphere lithography.[143] Two interesting examples are worth mentioning, NPG nanostructures prepared as nanodisks (as illustrated in Figure 1A) and plasmonic nanopore integrated in a NPG film. In the first case, NPG nanodisks demonstrated an SERS-EF up to $5 \times 10^8$ with excellent sensitivity with respect to not-patterned structures.[90] The second case, most recently demonstrated how NPG can be used as powerful platform for single molecule analysis in nanopore experiments, towards potential sequencing applications.[64,144]

On the other hand, enhanced UV spectroscopy has received keen interest because it offers interesting possibilities for studying electronic transition, selective molecular imaging, high resolution microscopy, as well as applications for photoelectric devices.[145] As most organic molecules feature strong absorption bands in the UV spectral domain[146], extending plasmonics into the UV range is of major interest for sensing and catalysis applications.[9,10,40,41,48,124,147–149] However, the commonly used plasmonic materials (*i.e.*, Au and Ag) support strong plasmon resonances only in the visible and NIR regimes, limiting plasmonic applications to these spectral ranges. To fill this gap, alternative plasmonic materials have been recently investigated. As already mentioned in previous section, aluminum, gallium, magnesium and rhodium are the most promising metals for this spectral range.[24] Interestingly, recently Dong *et al.* revisited the potential application of Si as a plasmonic material. Strong interband transitions lead to negative permittivity of Si across the ultraviolet spectral range, moreover, simulations demonstrated that Si nanodisk dimers can produce a local intensity enhancement greater than ~500-fold in a 1 nm gap at wavelength below 300 nm.[150]

Among the others, aluminum is the most investigated plasmonic material for the UV spectral regime.[22] Since Dorfer *et al.* demonstrated the SERS enhancement capabilities of thin aluminum layers for a deep-UV excitation wavelength of 244 nm,[151] various geometries of Al substrates have been investigated for the deep UV SERS.[38,152,153] Particularly, Mattiucci *et al.* theoretically demonstrated that, in spite of the fact that the ultraviolet SERS is limited by the metallic dampening, subwavelength Al gratings can yield as large SERS enhancement as $10^5$.[152] Aluminum film-over nanosphere (AlFON) substrates are also found to generate SERS enhancement factor of approximately $10^{(4-5)}$.[147] Similarly, Sigle *et al.* demonstrated that nanopatterned aluminum films with optimized surfaces (Figure 5B(a-c)) are capable to provide approximately 6 orders of magnitude SERS enhancement with deep-UV wavelength excitation ($\lambda_{excitation} = 244$ nm).[153] With respect to methods based on nanopatterning or nanoparticles synthesis, a nanoporous aluminum film can provide a low-cost platform where multiple LSPR hot-spots can be obtained. A major limitation with UV plasmonic metals, in particular with Al and Mg, is their reactivity with oxygen and consequent rapid surface oxidation.[154] This is true whit a thin film or a nanostructure of Al or Mg, and it is even more important in the typical dealloying processes to prepare NPMs. For this reason very few examples of NPMs made of Al and Mg have been reported so far.[9,10,50–53,155–157] Al and Mg are discussed here together because they have been used in combination or as starting alloy in order to obtain NPMs. In particular, Corsi *et al.* achieved a nanoporous Al structure starting from Al-Mg parent alloys made by melting pre Al and pure Mg at 750 °C. The nanoporous Al was fabricated by selective electrolytic removal of Mg from the parent alloy.[50] In their work, they demonstrated the use of this obtained NPM for the generation of hydrogen. A similar starting alloy was used by Garoli *et al.* to prepare nanoporous Al for plasmonic enhanced spectroscopies, in particular SERS (Figure 5B(d)) and fluorescence. In this case the procedure of fabrication was based on a Galvanic Replacement Reaction (GRR).[10,51] With respect to chemical dealloying, the GRR enables to prepare almost oxygen free porous films and, recently, this method is more and more explored. For example, Asselin *et al.* reported on the decoration of Mg nanoparticles to improve their plasmonic properties and Liu *et al.* reported on chirality transfer in NPMs during GRR.[158] It is extremely challenging to prepare nanoporous Al by means of chemical dealloying. Ponzellini *et al.* have recently obtained interesting results by using partially selective dealloying of Mg from a Mg-Al alloy. A metallic nanoporous film, with oxygen contents below 14% has been achieved limiting the dealloying duration in order to obtain a porous MgAl film with interesting plasmonic properties.[9] As for porous Al, also Mg is investigated as stand-alone porous metal.[53] It's now very interesting for plasmonic applications.[126] In fact, it can reversibly react with hydrogen to form magnesium hydride with consequent tunable optical response.

**Infrared Plasmonics and SPR-based Sensing**. Near-infrared spectroscopy can be used to obtain molecular and chemical information based on bands of the fundamental vibrational modes in the infrared wavelengths. Unfortunately, when compared to other wavelength ranges, the sensitivity of NIR spectroscopic measurement is limited by weak absorption and and consequent poor detection performances. To improve the performances of NIR spectroscopy, Shih and colleagues have developed a technique to simultaneously obtain chemical and refractive index sensing in 1− 2.5 μm NIR wavelength range using NPG disks where high-density plasmonic hot-spots can be generated.[92] Besides, Garoli *et al.* have carried out extensive investigations on the plasmonic properties, fabrication techniques, characterization, and functionalization of various architectures of NPG in the infrared regime for sensing applications.[11,70,71,88,89,159] Self-standing thin nanoporous gold leaves prepared by chemical dealloying of a silver-gold alloy film are found to have better reaction efficiency and detection sensitivity.[70] It has been demonstrated that a simple NPG film without complex nanofabrication process and optical setups can be used as a plasmonic metamaterial sensor characterized by a sensitivity up to 15,000 nm per RIU in the NIR range (Figure 6A).[89] This value is within the same order of magnitude as the state of the art metamaterials.[160] Similarly, 3D NPG resonators are found as promising sensing platforms in the near-infrared with sensitivity over 4000 nm/RIU,[161] leading to the detection of a small peptide (7-hystidine) with low concentrations (Figure 6B). The same 3D NPG resonators have been used to demonstrate the performance of NPG in Surface Enhanced Infrared Absorption (SEIRA).[162,159] In particular, by comparing the same 3D antenna prepared with bulk gold and NPG, it has been possible to verify a significant enhancement in the IR absorption from molecules adsorbed on the metal surface. The ultrahigh sensitivity of the NPG materials is attributed to the extreme sensitivity of surface plasmon resonances resulting from the change in the dielectric environment of metallic surfaces.[163] That is, when molecules are adsorbed to the surfaces of plasmonic nanostructure, the position of plasmon resonance of the nanostructure undergoes significant spectral shift.[127] This was extensively investigated also in the visible range where nanogrooves or nanoslits array of NPG demonstrated enhanced sensitivity with respect to similar structures prepared in bulk gold.[71,164] To evaluate the sensing performance of plasmonic sensors, the refractive index sensitivity (S) and the figure of merit (FoM) are often employed. The sensitivity of a plasmonic nanostructure to change in the local refractive index is usually expressed as the ratio between change in the spectral shift (Δλ) of the plasmon resonance of the nanostructure and the change in the refractive index ($\Delta n$) of the dielectric environment, *i.e.*, S = $\Delta\lambda/\Delta n$, thus expressed in terms of nanometer per refractive index unit (RIU). On the other hand, the figure of merit evaluates the precision of the sensing platform and it is often expressed as the ratio of sensitivity S and the full width at half maximum (FWHM) of the plasmon resonance spectra, *i.e.*, FOM = S/FWHM. [165]

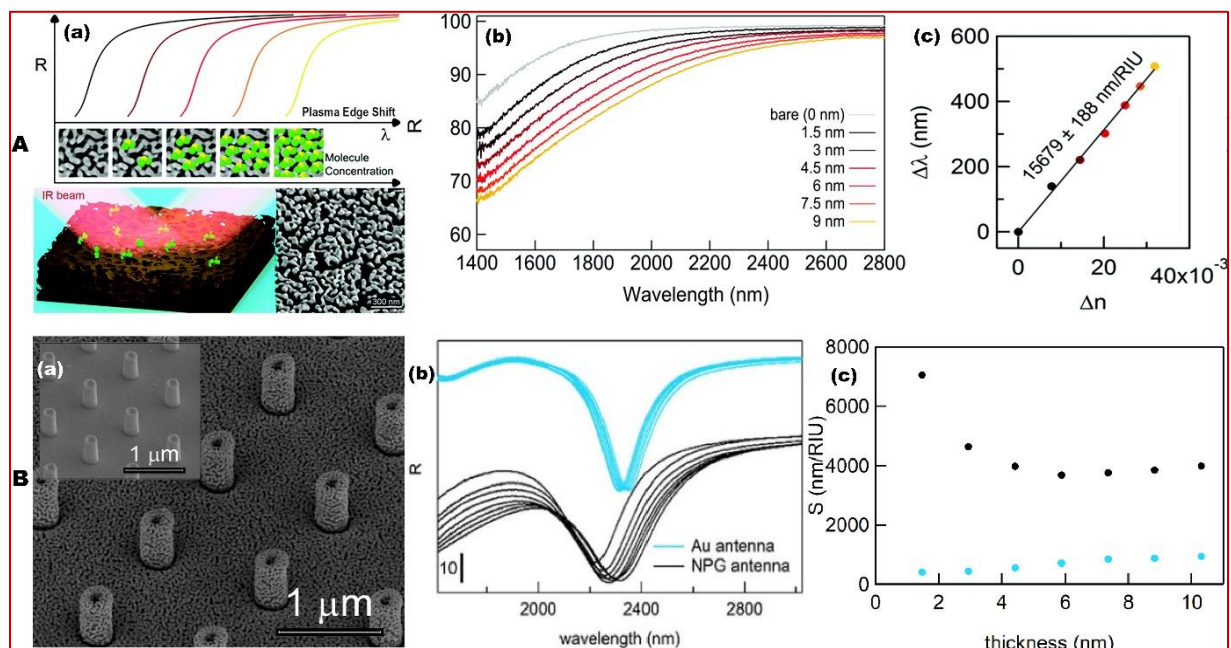

**Figure 6.** Nanoporous gold (NPG) platforms for high sensitivity IR plasmonic sensing. A) NPG metamaterials sensing performance with different surface coating. (a) Schematic of sensing approach: a near-infrared light beam illuminates the nanoporous gold, and the reflectance is measured around the plasma edge region where a significant spectral shift can be detected as a function of the number of molecules deposited on top of the material surface. (b) The reflectance curve of the NPG film as a function of the thickness of $SiO_2$ layer. (c) Corresponding spectral shift measured at R(0.85%). Reproduced with permission from ref. 89, Copyright 2019, The Royal Society of Chemistry. B) 3D NPG antenna for IR sensing. (a) SEM micrographs of the NPG vertical antenna. The inset displays title view of the same structure prepared in homogeneous gold (Au antenna). (b) Reflectance curves of 3D antenna arrays for NPG (black curves) and homogenous gold (blue curves). (c) Corresponding sensitivity of the NPG and homogenous Au antenna arrays. Reproduced with permission from ref. 161, Copyright 2019, Opt. Express.

**Metal-enhanced Fluorescence**. Plasmonic nanostructures have been widely exploited for the fluorescence emission enhancement, which has led to the widely explored field of the so called metal-enhanced fluorescence (MEF) or plasmon-enhanced fluorescence (PEF).[166] PEF has attracted enormous research interest [167] as it not only amplifies fluorescence emission intensity [168] but also provides the opportunity to perform imaging with resolutions significantly beyond the diffraction limit. [169] The plasmon-enhanced fluorescence benefits mainly from the local field enhancement (|E|) of the incident field and thus the fluorescence field enhancement is predicted to scale as square of the local field enhancement, *i.e.,* $|E|^2$ .[170] Thus, tailoring the porosity and surface morphology of NPM platforms is found to give rise to strong local field enhancement and ultrahigh fluorescence enhancement. [81,171,172] Generally, the fluorescence enhancement of NPMs relies not only on the near-field intensity but also on the distance between the metal and fluorophore as well as excitation wavelength .[13] As for the other plasmonic applications, also for MEF the most investigated NPM is NPG .[13,81,172–176] The effect of NPG on nearby fluorophores has been investigated considering well-defined metal-fluorophore distances. In this regard, Chen and colleagues [13] have investigated the distance effect by using silica as the spacing layer between fluorophores and NPG. The fluorescence enhancement of the silica coated NPG films was evaluated by using R6G fluorophore with the absorption and emission peaks at ~524 nm and ~553 nm, respectively, was chosen and stabilized on polymer substrate, SiO$_2$ coated polymer, NPG films and SiO$_2$@NPG films. However, it is evident from the Figure 7A that the fluorescence enhancement starts to drop for very short distances between the R6G and NPG film, indicating the onset of the quenching effect .[177] Theory predicts that fluorescence should drop to zero when the molecule is in direct contact with plasmonic particles.

Such quantitative studies on the plasmon-enhanced fluorescence have led to the development of metal-enhanced fluorescence imaging (Figure 7B), biosensing (Figure 7C) and therapeutic devices.[178] PEF bio applications techniques are primarily focused on substrate-based PEF and more recently has explored solution-based enhancement approaches. [179] Among several materials and architectures, NPG has been extensively explored for biosensing and imaging applications, owing to its physical properties such as excellent stability, biocompatibility, and high specific surface area to form self-assembled monolayers from thiols. [180,107,181] Particularly, Ahmed *et al.* have demonstrated a rapid, sensitive and quantitative detection of influenza A virus using a hybrid structure of quantum dot (QD) and nanoporous gold leaf (NPGL). [173] NPGL prepared by dealloying has well-defined surface morphology for binding anti-hemagglutinin (anti-HA) Ab-conjugated QDs (Ab-QDs) (Figure 7C). These bioconjugated components produce high PL intensity from QDs *via* surface plasmon resonance with the NPGL substrate, leading to three times higher PL intensity in the nanostructure of the antibody-functionalized NPGL than that without the NPGL for 10 μg/mL of HA. In the quantitative analysis using different concentrations of HA, PL intensities are found to be logarithmically dependent on the HA concentration, which is ranging from 1 ng/mL – 10 μg/mL (see Figure 7C (c) and the inset). The proposed PEF biosensing approach implies the potential of the platform for higher detection of much smaller volume of virus (up to 1 ng/mL) with less amount of reagents .[173]

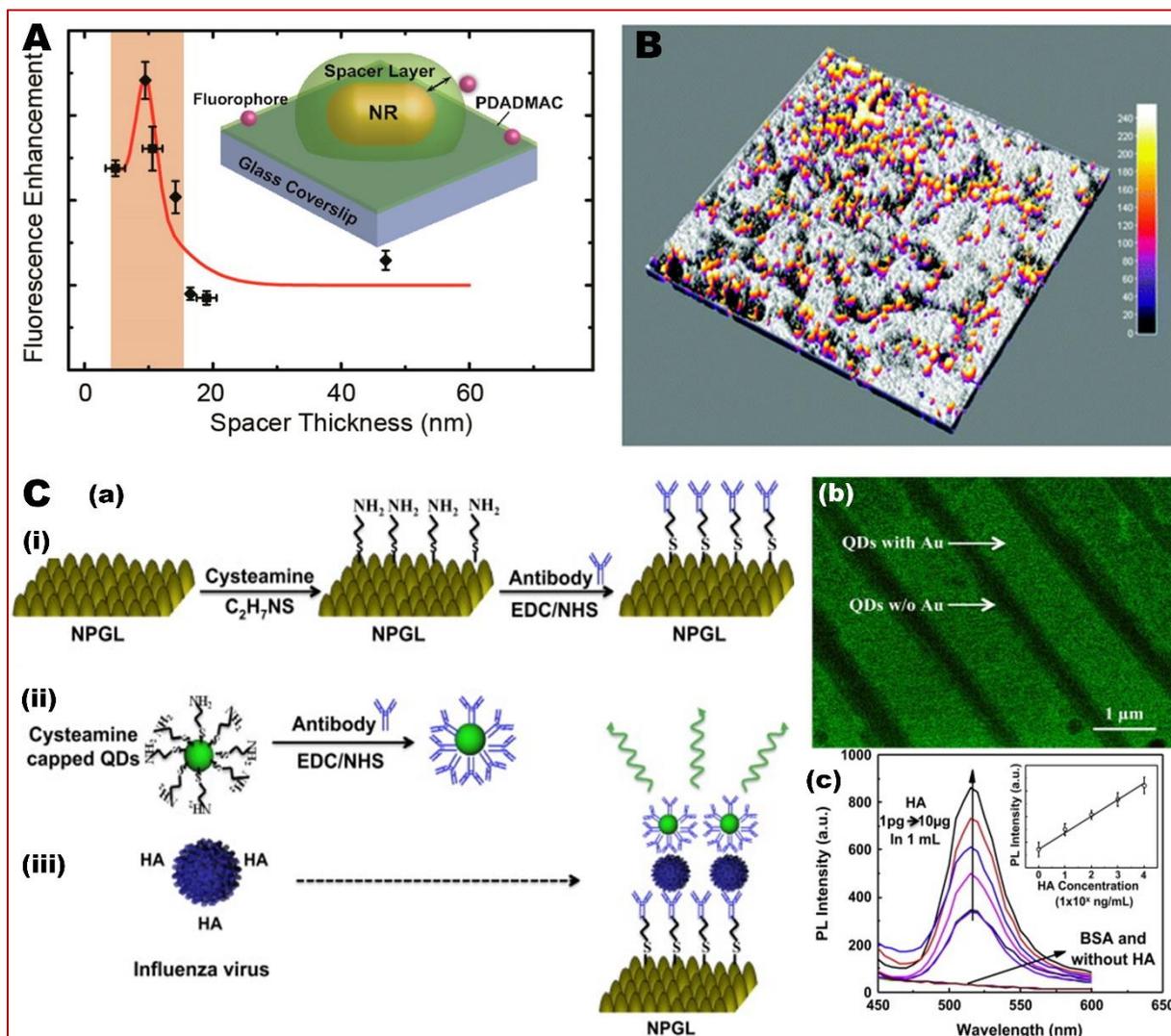

**Figure 7.** Nanoporous metal-based fluorescence emission enhancement, imaging and biosensing. A) Distance-dependent plasmon-enhanced fluorescence of single fluorescent molecules. Reproduced with permission from ref. 182, Copyright 2015, American Chemical Society. B) An overlaid image (40 × 40 $\mu m$) of fluorescence emission over a surface contour on a NPG film. The shaded contour represents surface feature derived from the reflectance image recorded from the same region. Reproduced with permission from ref. 172, Copyright 2013, The Royal Society of Chemistry. C) Nanoporous gold leaf (NPGL)-based assay for virus detection. (a) Schematic of the pandemic influenza virus A (H1N1) detection using hybrid structure of quantum dots (QDs) and nanoporous gold leaf. The NPGL (i) and QDs (ii) were initially conjugated with anti-hemagglutinin (HA) antibodies (anti-HA Ab, Y-shape) by the reaction of ethylcarbodiimide (EDC)/N-hydroxysuccinimide(NHS). Then anti-HA Ab-conjugated with NPGL and QDs form complex (iii) in presence of HA on the surface of influenza virus, finally enhancing PL intensity. (b) Fluorescence microscopic image of QDs on metallic nanostripe patterns. (c) PL enhancement corresponding to different quantities of recombinant influenza HA (H1N1) on anti- hemagglutinin (anti-HA) Ab-conjugated NPGL05. Inset: The calibration curve of PL intensity *versus* HA concentration. Reproduced with permission from ref. 173, Copyright 2014, Elsevier B.V.

NPMs can find interesting application also in MEF in the UV spectral region. As previously mentioned, metallic nanostructures for UV plasmonics are tipically prepared by means of electron beam lithography and focused ion beam lithography.[22] However, these fabrication processes require challenging optimization to achieve the very small nanostructures/nanogaps (5−10 nm) required for plasmonic resonances in the DUV region. Consequently these top-down techniques are not cost-effective and not recommended for large-area fabrication (cm$^2$).[183] Several alternative bottom-up approaches have been proposed, such as nanoimprint lithography,[184] electrochemical anodization,[185] and chemical synthesis of nanocrystals.[186,23] As previously mentioned, NPMs for UV have been investigated during the very recent years and their ability to enhance fluorescence in the UV has been reported, with an observed fluorescence enhancement up to 10 in the spectral range between 240 and 360 nm.[9,10]

**Extraordinary Light Transmission Effect and Fano Resonances**. Similar to localized plasmonic resonances surface plasmon polaritons can be created in NPM structures.[164,187,86] Kretschmann configuration is the conventional approach to excite SPPs propagating along the surface of semicontinuous metallic films.[188] Yu *et al.* has adapted this approach to achieve SPP generation in NPG films using a multilayer system.[86] In their experiments, they used light sources spanning over a broad wavelength range (594, 632.8, 780, 829 and 1152 nm) and demonstrated strong dependence of SPP generation on the excitation wavelength. At longer wavelengths (≥780 nm), a sharp reflection dip suggesting efficient SPP generation was shown. For wavelengths within the visible spectrum, SPR dip was broadened due to strong forward and directional backward scattering of SPPs as one would expect for nanoporous gold films with microscopic roughness.[189]. This observation is also consistent with other studies indicating the plasma frequency $\omega_p$ of nanoporous metals exhibit a spectral red-shift due to the lower density in comparison to bulk materials.[87]. A simpler approach to excite SPPs is to employ periodic perturbations defined on a NPM surface.[190]. Ruffato *et al.* exploited grating coupling mechanisms to excite SPPs on a NPM using directly incident light, enabling them to circumvent the need for prism coupling and the related alignment problems.[71] To fabricate periodically patterned NPG gratings, they used FIB lithography and employed a small molecule (Thiophenol, $C_6H_5SH$) surface functionalization scheme to compare NPG SPR sensitivity with respect to non-porous Au gratings. A greater resonance wavelength shift was observed in the case of patterned NPG film due to enhanced surface binding area; analyte molecules were able to penetrate into the pores and bind to the inner surfaces.

An interesting plasmonic phenomenon arises when the aforementioned periodic perturbations are defined in the form of subwavelength apertures (*e.g.*, nanoslits[191] and nanoholes[192]) perforating through a thin opaque metallic film. Light can be transmitted through these subwavelength apertures with orders of magnitude enhanced efficiencies than those of predicted by the Bethe's theory.[193,194] This so-called extraordinary transmission (EOT) effect has been used for a variety of applications (*e.g.,* microlenses, label-free biosensors and biochemical analysis).[195–201] The EOT effect in NPGs films was demonstratedusing nanoslits defined through NPG films.[164,202] To realize minimal damage to nanoporous structures during the FIB patterning process, they added a sacrificial Ag layer before the milling. The authors showed that this layer allows rapid heat dissipation and prevents melting of the porous gold structure. Following nanoslit patterning through the multilayer film, the sacrificial layer was readily removed using a selective etching process with $HNO_3$ at room temperature (20°C). Nanoslit devices with intact nanoporous structures were realized using this approach. The authors also showed four times increased sensitivity for molecular detection (DNA functionalization) with respect to nanoslit devices fabricated in non-porous gold films.

One interesting feature that arises in EOT devices is the Fano-resonant transmission profile, [192,203,204] an abrupt reversal of minimal to enhanced transmission within a narrow spectral window. This highly dispersive resonance transmission characteristic is a result of the intricate interactions between SPPs propagating along the surface and localized surface plasmons (LSPs) at the rims of the nanoapertures (Figure 8a-b).[205,206] Zhu *et al.* developed an intuitive phenomenological approach in order to explain the microscopic origins of the Fano resonant EOT effect .[207] For incident light to pass through subwavelength apertures efficiently, a series of excitation and coupling of plasmonic resonances (SPP → LSP → SPP) is needed. Using a coupled-oscillator model consists of mechanical analogues of the three plasmonic excitations responsible for EOT effect (Figure 8c-d), they were able to shed light on emergence of Fano-resonant behavior in EOT devices and biomolecular detection using mechanical loading concepts. They also demonstrated Fano-resonant EOT transmission in nanohole arrays (NHAs) fabricated using NPG (Figure 8e-f).[208] Presence of a spectrally sharp Fano resonance transmission profile in nanoporous NHAs indicate that strong interactions between SPPs and LSPs are preserved even in microscopically roughed surfaces. This observation is particularly important as it shows interacting plasmonic excitations can be created in NPM structures, giving rise to development of hybrid optical devices (*e.g.*, metasurfaces[209] and metamaterials[210,211]) utilizing multi-resonant behavior for efficient photon manipulation at the subwavelength scale.

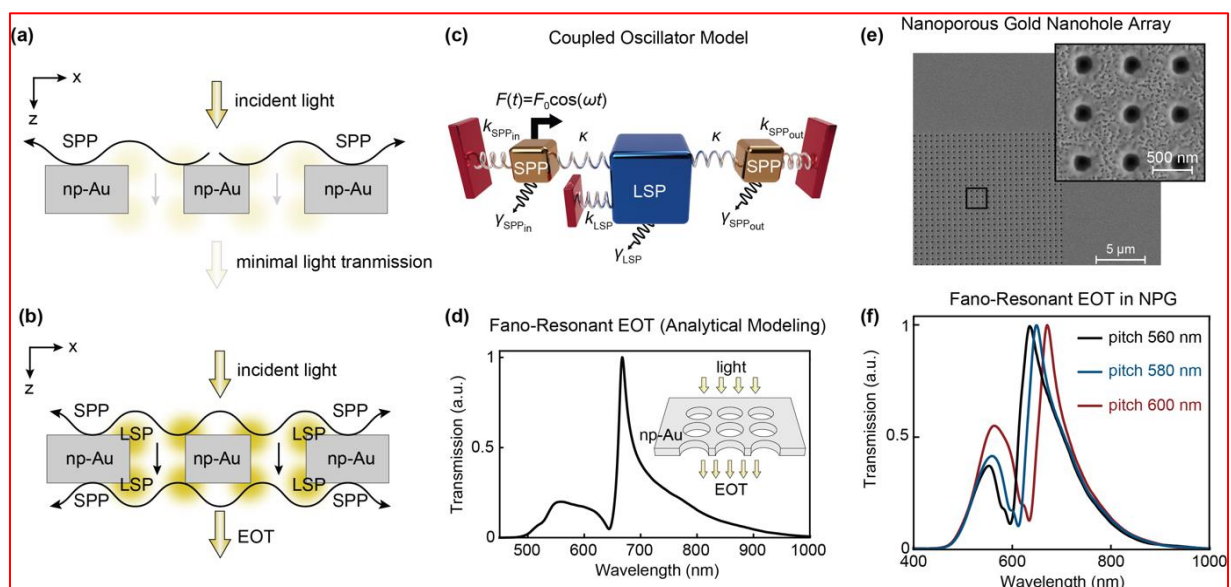

**Figure 8**. Extraordinary light transmission (EOT) effect and Fano resonances in nanoporous gold nanohole arrays. Schematic of SPP and LSP weak (a) and strong coupling (b) leading to transmission dip and peak is shown. (c) Coupled oscillator model, a mechanical analog of the three coupled plasmonic resonances, is shown. (d) Fano-resonant EOT spectra with highly dispersive character emerges from the oscillator model. (e) An SEM image of nanoporous plasmonic nanohole arrays are shown. Nanoporous features are significantly smaller with respect to nanohole openings. (f) Fano-resonant EOT transmission spectra obtained from broadband transmission measurements are shown. Adapted with permission from ref. 207, Copyright 2020, Elsevier B.V.

**Plasmon-enhanced Photocatalysis**
In the context of photocatalysis, metallic nanostructures with a characteristic length smaller than 30 nm can excite charge carriers upon LSPs decay.[58,57,212] These excited carriers, mostly hot-electrons,[213,57,58] may be transferred to molecules and other species in close proximity and drive chemical transformations.[214,215,216] Hence, NPMs can serve as platforms for photocatalysts.[214] In principle, the photocatalytic reactions taking place in the proximity of plasmonic nanostructures can depend on any of the energy conversion processes,[217,218] and their impact is generally difficult to completely untangle.[218] For example, plasmon-based high intensity near-fields are known to activate photosensitive reactions.[219] Additionally, such large near-fields can interact with nearby catalyzers, which can stimulate chemical reactions.[220] In this case, nanoparticles act as antennas that deliver electromagnetic energy to more chemically fit reactors in what have been named antenna-reactor systems.[221] Upon plasmon decay (~1-10 fs), both intra- and interband carriers can be excited, depending on the incident wavelength and the nanoparticle material. Both these carriers (electrons and holes) can transfer to nearby molecules if the electronic energy levels are properly aligned.[222] Moreover, intraband carriers can be further divided into 'hot' and 'thermalized' because they are typically generated from the relaxation of a plasmonic mode at different timescales (interband carriers are, instead, usually excited by non-plasmonic direct excitation).

Following plasmonic relaxation, energetic (hot) electron-hole pairs generate an out-of-equilibrium population, and its subsequent relaxation (~100s of fs) eventually constitutes the typical Fermi-Dirac electronic distribution. While in principle, both 'hot' and thermalized carriers can impact chemical reactions, usually one refers to hot-electron photochemistry to underline the plasmon relaxation phenomenon as the main driver of the process. On an even longer timescale (~10s of ps) electron-phonon scattering dominates, and the electronic distribution exchanges energy with the phononic lattice of the metals, thereby dissipating heat and increasing the nanoparticle temperature until an equilibrium is reached.[223] Finally, depending on the thermophysical properties of the hosting medium, the temperature of the nanoparticles equilibrates with the environment (~10s of ns). Chemical reactions typically follow an Arrhenius-like temperature dependence.[218] Therefore, plasmonic heating is one critical mechanism that is virtually always present and typically difficult to segregate from the other processes mentioned above. Moreover, while single/few plasmonic nanoparticles typically do not reach high temperatures unless irradiated by high-intensity focused light sources, arrays of nanoparticles exhibit collective heating effects, increasing the overall temperature even under mild illumination.[224] For these reasons, while plasmon-driven non-thermal catalytic effects have been demonstrated,[218] the impact of heating on plasmon-driven catalysis has sparked several discussions and debates.[225] A more detailed discussion of the current challenges and future perspectives of plasmon catalysis can be found in Ref.[218] Moreover, the effect of plasmonic excitations on more complex enzyme regulations has recently been discussed, highlighting the vast potential of light-nanoparticle interactions for both chemical and biological phenomena.[226]

As illustrated in Figure 9A, nanoporous metal-assisted photocatalysis takes place in the following mechanism. Upon illumination of nanoporous gold film, for example, surface plasmon modes are excited and local fields are enhanced on the surface of the metallic structure. As the plasmons decay, a hot carrier distribution is generated. Consequently electrons in the high-energy tail of this distribution can tunnel out of the metal into high-energy orbitals of the surrounding molecules catalysing the chemical reaction. Based on this principle, Wang and colleagues[217] have demonstrated the semi-conductor free photocatalytic efficiency of NPG structures. The bicontinuous NPG can be used to to enhance the electro-oxidation of alcohol molecules.[217] The NPG plasmonic catalyst enables to achieve the highest methanol oxidation current density of 531 $\mu A\ cm^{-2}$ among all known Au catalysts (Figure 9B). The higher hot carrier collection efficiency that can be achieved during direct plasmonic electrocatalysis without the assistance of Schottky junctions could impact in the development of high efficiency plasmonic catalysts for photo-enhanced electrochemical reactions. Similarly, taking advantage of visible light plasmonic-heating effect, Proschel et al.[218] also experimentally demonstrated significant enhancement of the kinetics of a redox reaction involving the oxidation of aluminum at the anode, and the reduction of hydrogen ions to hydrogen gas at the cathode (Figure 9C(a, b)). A 20-fold increase in the electrochemical current density upon exposure of the NPG cathode to visible light (Figure 9C(c)) is attributed to local heat generated in Au during localized surface plasmon resonance. Moreover, Ron and coworkers demonstrated a degradation of Rhodamine B molecules deposited on NPAg by visible light illumination.[59] Within 2 h thanks to NPAg, 75% of the Rhodamine B molecules were photocatalytically degraded as compared to 25% in the control experiment. The NPAg is able to absorb a large fraction of the solar spectrum and to generate energetic carriers. Further experiments demonstrated how the NPAg network can support high fraction of hot-carriers following a plasmonic decay.[59] For catalytic/photocatalytic applications, surface area is important, but high yield reactions can be achieved only with efficient molecular transport of both reactants and products. Better molecular transport can be achieved in multimodal pore sizes where large pores enable to easilyy transfer the molecules and to obtain sufficient diffusion rates. Moreover small pores serve as catalytic sites. Moreover, a limitation in photocatalysis applications of NPMs is the low stability of the materials to high temperatures. Biener *et al.* [227] demonstrated that atomic layer deposition (ALD) can be used to stabilize and functionalize nanoporous metals. Nanometer-thick alumina and titania ALD films can stabilize the nanoscale morphology of NPG up to 1000°C, with the additional effect of making the material stronger and stiffer. Moreover, the catalytic activity of NPG can be significantly increased by a coating of $TiO_2$ performed by means of ALD.

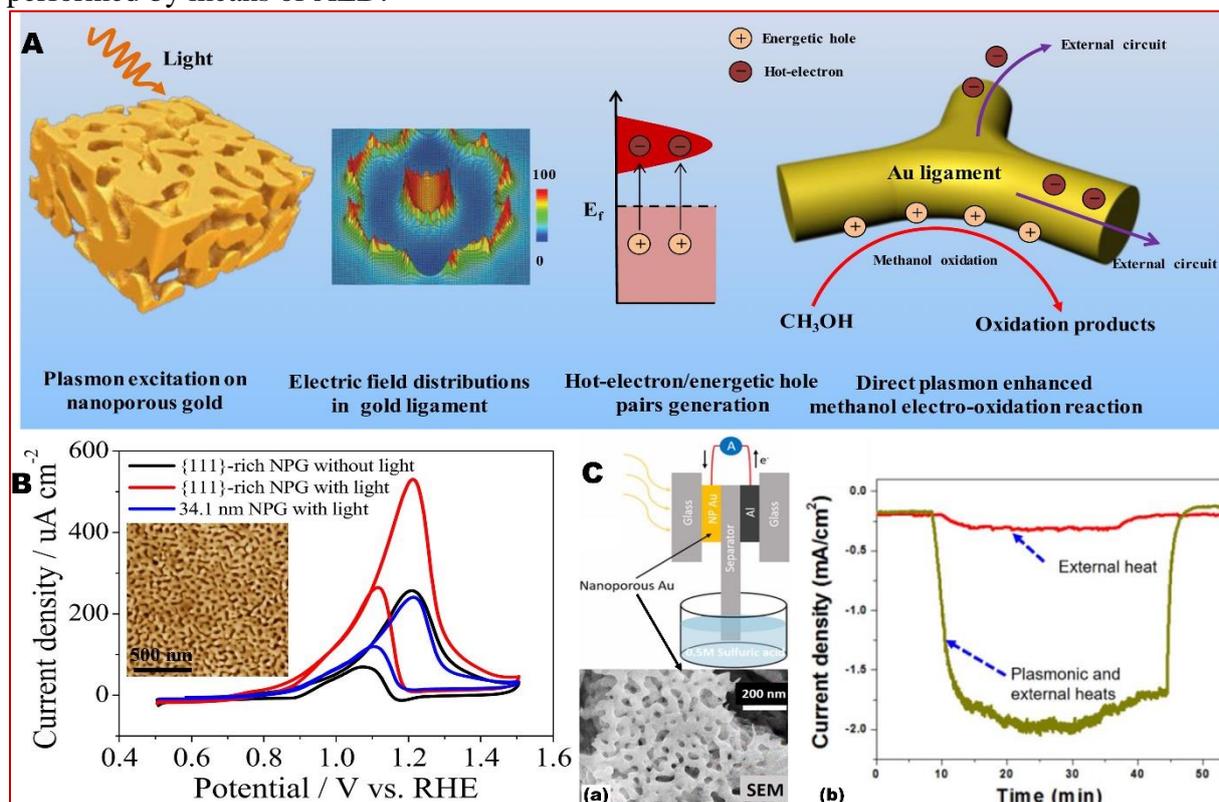

**Figure 9**. Plasmon-assisted photocatalysis based on nanoporous gold (NPG). A) Schematics of plasmon-induced hot electron/hole pairs in a NPG catalyst and the mechanistic representation of hole-assisted electro-oxidation of methanol on Au ligaments by pumping away hot electrons from the Schottky barrier-free plasmonic catalyst. B) Direct plasmon-enhanced electro-oxidation of methanol catalyzed by a surface-engineered NPG catalyst. The graph depics cyclic voltammetry curves of a methanol oxidation reaction on {111}-rich NPG and 34.1 nm NPG in a 0.5 M KOH/1.0 M methanol solution with and without light illumination (scan rate: 10 mV s$^{-1}$). The inset shows typical SEM image of NPG film. Adapted with permission from ref. 228, Copyright 2018, Elsevier Ltd. C) Plasmonic heating-enhanced electrochemical current in NPG cathodes. (a) Schematic of the photoelectrochemical current enhancement setup. The inset shows tpical scanning electron micrograph of NPG synthesized on glass slide substrate. (b) Comparison of the photoelectrochemical current density generated by external heat only (red), and by both plasmonic and external heats (green). Reproduced with permission under a Creative Commons Attribution 4.0 License (CC BY) from ref . 229, Copyright 2019, J. Electrochem. Soc.

**Conclusion, Challenges and Future Prospects**
In this critical review article, the fundamental plasmonic properties of nanoporous metals of various types and their functionalities for advanced spectroscopy and photocatalysis in broad range of spectra ranging from UV to mid-IR regimes are discussed. It is well noted that precise modelling and rational design of nanoporous metals play key roles in accurate prediction of the optical properties of the NPMs and, hence, enhance the performances of final devices. In this regard, thin films of NPMs with highly dense hotspots and widely tunable optical properties [7, 8, 27, 65, 66, 78, 79, 81] have received a great deal of interest for sensing, [18, 19] Raman enhancement [14, 21, 25, 140, 131, 230] and metal-enhanced fluorescence. [107] Moreover, given their structural features, strong polarization dependence, multiple resonance behavior, and ultrahigh local field enhancement,[231, 232] other nanoporous plasmonic nanostructures (such as nanoparticles, nanoantenna, and metamaterials) have also been widely explored for a variety of analytical and biomedical applications including single-particle SERS analysis, [233] IR plasmonic sensing, [89,159] and light-triggered drug delivery. [234]

Another important aspect of nanoporous metals is their fabrication strategy; with the dealloying technique that has been most widely utilized and facile method to fabricate nanoporous metals with three-dimensional bicontinuous porous configurations, open nanopores, tunable pore sizes. Unfortunately, it is not possible to perform dealloying from any starting alloy and the number of alloys suitable for the process is limited. Consequently the preparation of NPMs with desiderated properties is still challenging. Advanced approaches for dealloying are now under investigation, for example, metallic glasses can be used as precursors thanks to the good control that can achieved in composition. Moreover, if the dealloying process is combined with other strategies such as the templating, it could be possible to prepare NPMs with well controlled morphology and 3D structure. Finally, other methods such as electrochemical potential treatments on pure metals (such as porous Au and porous Cu electrodes), electrochemical deposition using $H_2$ bubbles as templates, metal nanoparticle self-assembly using a hydrothermal method, *etc* have been proposed for the preparation of NPMs and it is clear that improvement in these processes would certainly make NPMs more attractive in plasmonics.

Apart from accurate modelling, rational design and easy fabrication of nanoporous metals, material composition of NPMs also plays key role in determining their performances and functionalities. Although early works on nanoporous metals have been mainly focused on the traditional plasmonic metals (especially gold and silver), most recently, the use of other nanoporous metals such as Cu, Ni, Fe, Al and Rh. should be more promising for cost-effective sensing applications. Moreover, the most commonly utilized NPMs support plasmonic properties predominantly in the visible and infrared regimes of the electromagnetic spectrum. To fill this gap, alternative plasmonic materials such as Al, Mg, and Rh have emerged as the most promising metals for UV regimes. Specifically, owing to its large plasma frequency that leads to a negative permittivity (real part) down to the wavelength of 100 nm and strong local field enhancement because of its high electron density, Al has been a promising plasmonic material in the UV and DUV (DUV) regions. These interesting plasmonic properties of aluminum have resulted in numerous exciting applications including UV nanoantennas, DUV SERS, light emission enhancement of wide band gap semiconductors, improvement of light harvesting in solar cells, and UV metal-enhanced fluorescence.

Nanoporous metals with distinct three-dimensional bicontinous and interconnected porous configurations are promising materials as catalysts, fuel cells, supercapacitors, electrodes for batteries, and metamaterials. [59, 235, 26] However, pure nanoporous metallic devices suffer from poor conductivity and limited charge–discharge rate. This problem can be overcome by hybridizing transition-metal oxides with plasmonic materials. For instance, hybrid structures made of nanoporous gold and nanocrystalline $MnO_2$ can function as supercapacitors and electrodes for batteries, because of their high capacitance for storing electrical charge, leading to enhanced conductivity of the hybrid material, resulting in a specific capacitance close to the theoretical value of the constituent $MnO_2$ .[236]

Finally, while this paper discuss the plasmonic properties and applications of NPMs, the scientific community is still hardly working to improve the performances and to extend the field of applications of NPMs in general. Electrochemistry and the related field of energy systems, for example, are two important field of application of NPMs.[237] The advancements illustrated here on the preparation of non-noble porous metals could be applied to electrochemical sensors and devices. The potential use of bi-metallic NPMs could introduce additional functionalities in a close future not only in plasmonics. Porous metals can for example coupled with metal oxides, conducting polymers and alloys to enhance electron and ion mobility. With the continuous advancement in fabrication and characterization we can expect several years of technical improvement and cutting-edge scientific results in the field of porous metal and porous materials.

**Funding**

The research leading to these results has received funding from the European Union under the Marie Skłodowska-Curie RISE project COMPASS No. 691185, by the National Science Foundation under Grant No. (IIP-1941227) and by Horizon 2020 Program, FET-Open: DNA-FAIRYLIGHTS, Grant Agreement no. 964995.

**Vocabulary**

**Nanoporous metals**, artificially designed metamaterials made of solid metals with nanosized porosity, ultrahigh specific surface area, good electrical conductivity, high structural stability, and tunable optical properties; **dealloying**, the selective leaching of one or more components out of a solid solution alloy to produce a residual nanoporous structure; **surface plasmon resonance**, a resonant oscillation of conduction band free electrons in plasmonic nanostructures; **plasmon-enhanced spectroscopy**, an advanced spectroscopic technique that exploits the potentials of plasmonic effects such as surface plasmon resonance and local field enhancement to amplify, for example, the Raman signal and fluorescence emission of small molecules; **plasmonic photocatalysis**, an accelerated and enhanced photocatalysis process through the effect of surface plasmon resonance.

**References**
(1) Koya, A. N.; Cunha, J.; Guo, T. L.; Toma, A.; Garoli, D.; Wang, T.; Juodkazis, S.; Cojoc, D.; Proietti Zaccaria, R. Novel Plasmonic Nanocavities for Optical Trapping-Assisted Biosensing Applications. *Adv. Opt. Mater.* **2020**, *8* (7).
(2) Fujita, T. Hierarchical Nanoporous Metals as a Path toward the Ultimate Three-Dimensional Functionality. *Sci. Technol. Adv. Mater.* **2017**, *18* (1), 724–740.
(3) Qiu, H. J.; Li, X.; Xu, H. T.; Zhang, H. J.; Wang, Y. Nanoporous Metal as a Platform for Electrochemical and Optical Sensing. *J. Mater. Chem. C* **2014**, *2* (46), 9788–9799.
(4) Qiu, H.-J.; Xu, H.-T.; Liu, L.; Wang, Y. Correlation of the Structure and Applications of Dealloyed Nanoporous Metals in Catalysis and Energy Conversion/Storage. *Nanoscale* **2015**, *7* (2), 386–400.
(5) Ding, Y.; Zhang, Z. *Nanoporous Metals for Advanced Energy Technologies*; Springer International Publishing: Cham, 2016.
(6) Koya, A. N.; Lin, J. Charge Transfer Plasmons: Recent Theoretical and Experimental Developments. *Appl. Phys. Rev.* **2017**, *4* (2), 021104.
(7) Lang, X.; Qian, L.; Guan, P.; Zi, J.; Chen, M. Localized Surface Plasmon Resonance of Nanoporous Gold. *Appl. Phys. Lett.* **2011**, *98* (9), 3–5.
(8) Detsi, E.; Salverda, M.; Onck, P. R.; De Hosson, J. T. M. On the Localized Surface Plasmon Resonance Modes in Nanoporous Gold Films. *J. Appl. Phys.* **2014**, *115* (4).
(9) Ponzellini, P.; Giovannini, G.; Cattarin, S.; Zaccaria, P.R.; Marras, S.; Prato, M.; Schirato, A.; D'Amico, F.; Calandrini, E.; De Angelis, F.; Yang, W.; Jin, H.-J.; Alabastri, A.; Garoli, D. Metallic Nanoporous Aluminum-Magnesium Alloy for UV-Enhanced Spectroscopy. *J. Phys. Chem. C* **2019**, *123* (33), 20287–20296.
(10) 10. Garoli, D.; Schirato, A.; Giovannini, G.; Cattarin, S.; Ponzellini, P.; Calandrini, E.; Zaccaria, R. P.; D'Amico, F.; Pachetti, M.; Yang, W.; Jin, H.-J.; Krahne, R.; Alabastri,

A. Galvanic Replacement Reaction as a Route to Prepare Nanoporous Aluminum for UV Plasmonics. *Nanomaterials* **2020**, *10* (1), 1–12.

(11) Garoli, D.; Calandrini, E.; Bozzola, A.; Toma, A.; Cattarin, S.; Ortolani, M.; De Angelis, F. Fractal-Like Plasmonic Metamaterial with a Tailorable Plasma Frequency in the Near-Infrared. *ACS Photonics* **2018**, *5* (8), 3408–3414.

(12) Liu, K.; Bai, Y.; Zhang, L.; Yang, Z.; Fan, Q.; Zheng, H.; Yin, Y.; Gao, C. Porous Au–Ag Nanospheres with High-Density and Highly Accessible Hotspots for SERS Analysis. *Nano Lett.* **2016**, *16* (6), 3675–3681.

(13) Chen, C.; Zhang, L.; Yang, M.; Tao, C.; Han, Z.; Chen, B.; Zeng, H. Size and Distance Dependent Fluorescence Enhancement of Nanoporous Gold. *Opt. Express* **2017**, *25* (9), 9901.

(14) Qian, L. H.; Yan, X. Q.; Fujita, T.; Inoue, A.; Chen, M. W. Surface Enhanced Raman Scattering of Nanoporous Gold: Smaller Pore Sizes Stronger Enhancements. *Appl. Phys. Lett.* **2007**, *90* (15), 3–6.

(15) Lang, X. Y.; Chen, L. Y.; Guan, P. F.; Fujita, T.; Chen, M. W. Geometric Effect on Surface Enhanced Raman Scattering of Nanoporous Gold: Improving Raman Scattering by Tailoring Ligament and Nanopore Ratios. *Appl. Phys. Lett.* **2009**, *94* (21), 2009–2011.

(16) Jiao, Y.; Ryckman, J. D.; Ciesielski, P. N.; Escobar, C. A.; Jennings, G. K.; Weiss, S. M. Patterned Nanoporous Gold as an Effective SERS Template. *Nanotechnology* **2011**, *22* (29).

(17) Qiu, H.; Zhang, Z.; Huang, X.; Qu, Y. Dealloying Ag-Al Alloy to Prepare Nanoporous Silver as a Substrate for Surface-Enhanced Raman Scattering: Effects of Structural Evolution and Surface Modification. *ChemPhysChem* **2011**, *12* (11), 2118–2123.

(18) Ma, C.; Trujillo, M. J.; Camden, J. P. Nanoporous Silver Film Fabricated by Oxygen Plasma: A Facile Approach for SERS Substrates. *ACS Appl. Mater. Interfaces* **2016**, *8* (36), 23978–23984.

(19) Jiang, R.; Xu, W.; Wang, Y.; Yu, S. Tunable Porous Silver Nanostructures for Efficient Surface-Enhanced Raman Scattering Detection of Trace Pesticide Residues. *New J. Chem.* **2018**, *42* (21), 17750–17755.

(20) Chen, L. Y.; Yu, J. S.; Fujita, T.; Chen, M. W. Nanoporous Copper with Tunable Nanoporosity for SERS Applications. *Adv. Funct. Mater.* **2009**, *19* (8), 1221–1226.

(21) Song, R.; Zhang, L.; Zhu, F.; Li, W.; Fu, Z.; Chen, B.; Chen, M.; Zeng, H.; Pan, D. Hierarchical Nanoporous Copper Fabricated by One-Step Dealloying Toward Ultrasensitive Surface-Enhanced Raman Sensing. *Adv. Mater. Interfaces* **2018**, *5* (16), 1800332.

(22) Knight, M. W.; King, N. S.; Liu, L.; Everitt, H. O.; Nordlander, P.; Halas, N. J. Aluminum for Plasmonics. *ACS Nano* **2014**, *8* (1), 834–840.

(23) Watson, A. M.; Zhang, X.; Alcaraz De La Osa, R.; Sanz, J. M.; González, F.; Moreno, F.; Finkelstein, G.; Liu, J.; Everitt, H. O. Rhodium Nanoparticles for Ultraviolet Plasmonics. *Nano Lett.* **2015**, *15* (2), 1095–1100.

(24) Gutiérrez, Yael; Alcaraz de la Osa, Rodrigo; Ortiz, Dolores ; Saiz, José María; González, Francisco; Moreno, F. Plasmonics in the Ultraviolet with Aluminum, Gallium, Magnesium and Rhodium. *Appl. Sci.* **2018**, *8* (1), 64.

(25) Zhang, L.; Lang, X.; Hirata, A.; Chen, M. Wrinkled Nanoporous Gold Films with Ultrahigh Surface-Enhanced Raman Scattering Enhancement. *ACS Nano* **2011**, *5* (6), 4407–4413.

(26) Ron, R.; Haleva, E.; Salomon, A. Nanoporous Metallic Networks: Fabrication, Optical Properties, and Applications. *Adv. Mater.* **2018**, *30* (41), 1–14.

(27) Jalas, D.; Canchi, R.; Petrov, A. Y.; Lang, S.; Shao, L.; Weissmüller, J.; Eich, M. Effective Medium Model for the Spectral Properties of Nanoporous Gold in the

Visible. *Appl. Phys. Lett.* **2014**, *105* (24), 241906.

(28) Liu, G.; Li, K.; Zhang, Y.; Du, J.; Ghafoor, S.; Lu, Y. A Facile Periodic Porous Au Nanoparticle Array with High-Density and Built-in Hotspots for SERS Analysis. *Appl. Surf. Sci.* **2020**, *527*, 146807.

(29) Daggumati, P.; Matharu, Z.; Seker, E. Effect of Nanoporous Gold Thin Film Morphology on Electrochemical DNA Sensing. *Anal. Chem.* **2015**, *87* (16), 8149–8156.

(30) Qiu, S.; Zhao, F.; Zenasni, O.; Li, J.; Shih, W.-C. Nanoporous Gold Disks Functionalized with Stabilized G-Quadruplex Moieties for Sensing Small Molecules. *ACS Appl. Mater. Interfaces* **2016**, *8* (44), 29968–29976.

(31) Rebbecchi, T. A.; Chen, Y. Template-Based Fabrication of Nanoporous Metals. *J. Mater. Res.* **2018**, *33* (1), 2–15.

(32) Weissmüller, J.; Newman, R. C.; Jin, H.-J.; Hodge, A. M.; Kysar, J. W. Nanoporous Metals by Alloy Corrosion: Formation and Mechanical Properties. *MRS Bull.* **2009**, *34* (8), 577–586.

(33) Zhao, F.; Zeng, J.; Arnob, Md M. P.; Sun, P.; Qi, J.; Motwani, P.; Gheewala, M.; Li, C.-H.; Paterson, A.; Strych, U.; Raja, B.; Willson, R.C.; Wolfe, J. C.; Leeb, T. R.; Shih, W.-C. Monolithic NPG Nanoparticles with Large Surface Area, Tunable Plasmonics, and High-Density Internal Hot-Spots. *Nanoscale* **2014**, *6* (14), 8199–8207.

(34) Arnob, M. M. P.; Artur, C.; Misbah, I.; Mubeen, S.; Shih, W. C. 10×-Enhanced Heterogeneous Nanocatalysis on a Nanoporous Gold Disk Array with High-Density Hot Spots. *ACS Appl. Mater. Interfaces* **2019**, *11* (14), 13499–13506.

(35) Seok, J. Y.; Lee, J.; Yang, M. Self-Generated Nanoporous Silver Framework for High-Performance Iron Oxide Pseudocapacitor Anodes. *ACS Appl. Mater. Interfaces* **2018**, *10* (20), 17223–17231.

(36) Shen, Z.; O'Carroll, D. M. Nanoporous Silver Thin Films: Multifunctional Platforms for Influencing Chain Morphology and Optical Properties of Conjugated Polymers. *Adv. Funct. Mater.* **2015**, *25* (22), 3302–3313.

(37) Chowdhury, M. H.; Ray, K.; Gray, S. K.; Pond, J.; Lakowicz, J. R. Aluminum Nanoparticles as Substrates for Metal-Enhanced Fluorescence in the Ultraviolet for the Label-Free Detection of Biomolecules. *Anal. Chem.* **2009**, *81* (4), 1397–1403.

(38) Jha, S. K.; Ahmed, Z.; Agio, M.; Ekinci, Y.; Löffler, J. F. Deep-UV Surface-Enhanced Resonance Raman Scattering of Adenine on Aluminum Nanoparticle Arrays. *J. Am. Chem. Soc.* **2012**, *134* (4), 1966–1969.

(39) McPeak, K. M.; van Engers, C.D.; Bianchi, S.; Rossinelli, A.; Poulikakos, L. V.; Bernard, L.; Hermann, S.; Kim, D. K.; Burger, S.; Blome, M.; Jayanti, S.V.; Norris, D. J. Ultraviolet Plasmonic Chirality from Colloidal Aluminum Nanoparticles Exhibiting Charge-Selective Protein Detection. *Adv. Mater.* **2015**, *27* (40), 6244–6250.

(40) Jiao, X.; Peterson, E. M.; Harris, J. M.; Blair, S. UV Fluorescence Lifetime Modification by Aluminum Nanoapertures. *ACS Photonics* **2014**, *1* (12), 1270–1277.

(41) Sharma, B.; Cardinal, M. F.; Ross, M. B.; Zrimsek, A. B.; Bykov, S. V.; Punihaole, D.; Asher, S. A.; Schatz, G. C.; Van Duyne, R. P. Aluminum Film-Over-Nanosphere Substrates for Deep-UV Surface-Enhanced Resonance Raman Spectroscopy. *Nano Lett.* **2016**, *16* (12), 7968–7973.

(42) Ray, K.; Chowdhury, M. H.; Lakowicz, J. R. Aluminum Nanostructured Films as Substrates for Enhanced Fluorescence in the Ultraviolet-Blue Spectral Region. *Anal. Chem.* **2007**, *79* (17), 6480–6487.

(43) Hao, Q.; Wang, C.; Huang, H.; Li, W.; Du, D.; Han, D.; Qiu, T.; Chu, P. K. Aluminum Plasmonic Photocatalysis. *Sci. Rep.* **2015**, *5*, 1–7.

(44) Chou, B. T.; Chou, Y. H.; Wu, Y. M.; Chung, Y. C.; Hsueh, W. J.; Lin, S. W.; Lu, T. C.; Lin, T. R.; Lin, S. Di. Single-Crystalline Aluminum Film for Ultraviolet Plasmonic

Nanolasers. *Sci. Rep.* **2016**, *6* (December 2015), 1–9.

(45) Ekinci, Y.; Solak, H. H.; Löffler, J. F. Plasmon Resonances of Aluminum Nanoparticles and Nanorods. *J. Appl. Phys.* **2008**, *104* (8).

(46) Tian, S.; Neumann, O.; McClain, M. J.; Yang, X.; Zhou, L.; Zhang, C.; Nordlander, P.; Halas, N. J. Aluminum Nanocrystals: A Sustainable Substrate for Quantitative SERS-Based DNA Detection. *Nano Lett.* **2017**, *17* (8), 5071–5077.

(47) Zhang, F.; Martin, J.; Plain, J. Long-Term Stability of Plasmonic Resonances Sustained by Evaporated Aluminum Nanostructures. *Opt. Mater. Express* **2019**, *9* (1), 85.

(48) Norek, M.; Włodarski, M.; Matysik, P. UV Plasmonic-Based Sensing Properties of Aluminum Nanoconcave Arrays. *Curr. Appl. Phys.* **2014**, *14* (11), 1514–1520.

(49) Martin, J.; Proust, J.; Gérard, D.; Plain, J. Localized Surface Plasmon Resonances in the Ultraviolet from Large Scale Nanostructured Aluminum Films. *Opt. Mater. Express* **2013**, *3* (7), 954.

(50) Corsi, J. S.; Fu, J.; Wang, Z.; Lee, T.; Ng, A. K.; Detsi, E. Hierarchical Bulk Nanoporous Aluminum for On-Site Generation of Hydrogen by Hydrolysis in Pure Water and Combustion of Solid Fuels. *ACS Sustain. Chem. Eng.* **2019**, *7*, 11194–11204.

(51) Yang, W.; Zheng, X.-G.; Wang, S.-G.; Jin, H.-J. Nanoporous Aluminum by Galvanic Replacement: Dealloying and Inward-Growth Plating. *J. Electrochem. Soc.* **2018**, *165* (9), C492–C496.

(52) Soler, R.; Suárez, O. M.; Declet, A.; Hernández-Maldonado, A. J.; Estremera, E. G. Fabrication of Porous and Nanoporous Aluminum *via* Selective Dissolution of Al-Zn Alloys. *Adv. Mater. Sci. Eng.* **2014**, *2014*, 1–6.

(53) Liu, J.; Wang, H.; Yuan, Q.; Song, X. A Novel Material of Nanoporous Magnesium for Hydrogen Generation with Salt Water. *J. Power Sources* **2018**, *395* (April), 8–15.

(54) Ahmadivand, A.; Sinha, R.; Kaya, S.; Pala, N. Rhodium Plasmonics for Deep-Ultraviolet Bio-Chemical Sensing. *Plasmonics* **2016**, *11* (3), 839–849.

(55) Jiang, B.; Li, C.; Dag, Ö.; Abe, H.; Takei, T.; Imai, T.; Hossain, Md. S. A.; Islam, Md. T.; Wood, K.; Henzie, J.; Yamauchi, Y. Mesoporous Metallic Rhodium Nanoparticles. *Nat. Commun.* **2017**, *8* (May), 3–4.

(56) Zhang, X.; Li, P.; Barreda, Á.; Gutiérrez, Y.; González, F.; Moreno, F.; Everitt, H. O.; Liu, J. Size-Tunable Rhodium Nanostructures for Wavelength-Tunable Ultraviolet Plasmonics. *Nanoscale Horizons* **2016**, *1* (1), 75–80.

(57) Brongersma, M. L.; Halas, N. J.; Nordlander, P. Plasmon-Induced Hot Carrier Science and Technology. *Nat. Nanotechnol.* **2015**, *10* (1), 25–34.

(58) Clavero, C. Plasmon-Induced Hot-Electron Generation at Nanoparticle/Metal-Oxide Interfaces for Photovoltaic and Photocatalytic Devices. *Nat. Photonics* **2014**, *8* (2), 95–103.

(59) Ron, R.; Gachet, D.; Rechav, K.; Salomon, A. Direct Fabrication of 3D Metallic Networks and Their Performance. *Adv. Mater.* **2017**, *29* (7), 1604018.

(60) Ron, R.; Shavit, O.; Aharon, H.; Zielinski, M.; Galanty, M.; Salomon, A. Nanoporous Metallic Network as a Large-Scale 3D Source of Second Harmonic Light. *J. Phys. Chem. C* **2019**, *123* (41), 25331–25340.

(61) Xia, R.; Wu, R. N.; Liu, Y. L.; Sun, X. Y. The Role of Computer Simulation in Nanoporous Metals-a Review. *Materials (Basel).* **2015**, *8* (8), 5060–5083.

(62) Yildiz, Y. O.; Kirca, M. Atomistic Simulation of Voronoi-Based Coated Nanoporous Metals. *Model. Simul. Mater. Sci. Eng.* **2017**, *25* (2), 025008.

(63) Wang, J.; Lam, D. C. C. Model and Analysis of Size-Stiffening in Nanoporous Cellular Solids. *J. Mater. Sci.* **2009**, *44* (4), 985–991.

(64) Hubarevich, A.; Huang, J.-A.; Giovannini, G.; Schirato, A.; Zhao, Y.; Maccaferri, N.; De Angelis, F.; Alabastri, A.; Garoli, D. λ-DNA through Porous Materials—Surface-Enhanced Raman Scattering in a Simple Plasmonic Nanopore. *J. Phys. Chem. C* **2020**, *124* (41), 22663–22670.

(65) Zhou, F.; Deng, Q.; Li, X.; Shao, L. H. Investigation of the Distinct Optical Property of Nanoporous Gold. *Results Phys.* **2019**, *15* (September), 102645.

(66) Jalas, D.; Shao, L. H.; Canchi, R.; Okuma, T.; Lang, S.; Petrov, A.; Weissmüller, J.; Eich, M. Electrochemical Tuning of the Optical Properties of Nanoporous Gold. *Sci. Rep.* **2017**, *7* (September 2016), 1–8.

(67) Ramesh, R.; Niauzorau, S.; Ni, Q.; Azeredo, B. P.; Wang, L. Optical Characterization and Modeling of Nanoporous Gold Absorbers Fabricated by Thin-Film Dealloying. *Nanotechnology* **2020**, *31* (40), 405706.

(68) Gehr, R. J.; Boyd, R. W. Optical Properties of Nanostructured Optical Materials. *Chem. Mater.* **1996**, *8* (8), 1807–1819.

(69) Lifshitz, D. L. a. E. M. L. *Electrodynamics of Continous Media*, Pergamon P.; 1960.

(70) Garoli, D.; Ruffato, G.; Zilio, P.; Calandrini, E.; De Angelis, F.; Romanato, F.; Cattarin, S. Nanoporous Gold Leaves: Preparation, Optical Characterization and Plasmonic Behavior in the Visible and Mid-Infrared Spectral Regions. *Opt. Mater. Express* **2015**, *5* (10).

(71) Ruffato, G.; Romanato, F.; Garoli, D.; Cattarin, S. Nanoporous Gold Plasmonic Structures for Sensing Applications. *Opt. Express* **2011**, *19* (14), 13164–13170.

(72) Rao, W.; Wang, D.; Kups, T.; Baradács, E.; Parditka, B.; Erdélyi, Z.; Schaaf, P. Nanoporous Gold Nanoparticles and Au/Al2O3 Hybrid Nanoparticles with Large Tunability of Plasmonic Properties. *ACS Appl. Mater. Interfaces* **2017**, *9* (7), 6273–6281.

(73) Arnold, M. D.; Blaber, M. G. Optical Performance and Metallic Absorption in Nanoplasmonic Systems. *Opt. Express* **2009**, *17* (5), 3835.

(74) Tian, Z. Q.; Ren, B.; Wu, D. Y. Surface-Enhanced Raman Scattering: From Noble to Transition Metals and from Rough Surfaces to Ordered Nanostructures. *J. Phys. Chem. B* **2002**, *106* (37), 9463–9483.

(75) Jena, B. K.; Raj, C. R. Seedless, Surfactantless Room Temperature Synthesis of Single Crystalline Fluorescent Gold Nanoflowers with Pronounced SERS and Electrocatalytic Activity. *Chem. Mater.* **2008**, *20* (11), 3546–3548.

(76) Ortolani, M.; Mancini, A.; Budweg, A.; Garoli, D.; Brida, D.; de Angelis, F. Pump-Probe Spectroscopy Study of Ultrafast Temperature Dynamics in Nanoporous Gold. *Phys. Rev. B* **2019**, *99* (3), 035435.

(77) Stine, K. J. Nanoporous Gold and Other Related Materials. *Nanomaterials* **2019**, *9* (8), 13–15.

(78) Chen, H. A.; Long, J. L.; Lin, Y. H.; Weng, C. J.; Lin, H. N. Plasmonic Properties of a Nanoporous Gold Film Investigated by Far-Field and Near-Field Optical Techniques. *J. Appl. Phys.* **2011**, *110* (5), 0–5.

(79) Qian, L.; Shen, B.; Qin, G. W.; Das, B. Widely Tuning Optical Properties of Nanoporous Gold-Titania Core-Shells. *J. Chem. Phys.* **2011**, *134* (1), 014707.

(80) Qian, L.; Das, B.; Li, Y.; Yang, Z. Giant Raman Enhancement on Nanoporous Gold Film by Conjugating with Nanoparticles for Single-Molecule Detection. *J. Mater. Chem.* **2010**, *20* (33), 6891–6895.

(81) Lang, X. Y.; Guan, P. F.; Zhang, L.; Fujita, T.; Chen, M. W. Size Dependence of Molecular Fluorescence Enhancement of Nanoporous Gold. *Appl. Phys. Lett.* **2010**, *96* (7), 16–19.

(82) Ding, Y.; Kim, Y. J.; Erlebacher, J. Nanoporous Gold Leaf: "Ancient Technology "/Advanced Material. *Adv. Mater.* **2004**, *16* (21), 1897–1900.

(83) Bhattarai, J. K.; Neupane, D.; Nepal, B.; Mikhaylov, V.; Demchenko, A. V.; Stine, K. J. Preparation, Modification, Characterization, and Biosensing Application of Nanoporous Gold Using Electrochemical Techniques. *Nanomaterials* **2018**, *8* (3).

(84) Seker, E.; Reed, M. L.; Begley, M. R. Nanoporous Gold: Fabrication, Characterization, and Applications. *Materials (Basel).* **2009**, *2* (4), 2188–2215.

(85) Biener, J.; Nyce, G. W.; Hodge, A. M.; Biener, M. M.; Hamza, A. V.; Maier, S. A. Nanoporous Plasmonic Metamaterials. *Adv. Mater.* **2008**, *20* (6), 1211–1217.

(86) Yu, F.; Ahl, S.; Caminade, A. M.; Majoral, J. P.; Knoll, W.; Erlebacher, J. Simultaneous Excitation of Propagating and Localized Surface Plasmon Resonance in Nanoporous Gold Membranes. *Anal. Chem.* **2006**, *78* (20), 7346–7350.

(87) Maaroof, A. I.; Gentle, A.; Smith, G. B.; Cortie, M. B. Bulk and Surface Plasmons in Highly Nanoporous Gold Films. *J. Phys. D. Appl. Phys.* **2007**, *40* (18), 5675–5682.

(88) Ruffato, G.; Garoli, D.; Cattarin, S.; Barison, S.; Natali, M.; Canton, P.; Benedetti, A.; De Salvador, D.; Romanato, F. Patterned Nanoporous-Gold Thin Layers: Structure Control and Tailoring of Plasmonic Properties. *Microporous Mesoporous Mater.* **2012**, *163*, 153–159.

(89) Garoli, D.; Calandrini, E.; Giovannini, G.; Hubarevich, A.; Caligiuri, V.; De Angelis, F. Nanoporous Gold Metamaterials for High Sensitivity Plasmonic Sensing. *Nanoscale Horizons* **2019**, *4* (5), 1153–1157.

(90) Qi, J.; Motwani, P.; Gheewala, M.; Brennan, C.; Wolfe, J. C.; Shih, W.-C. Surface-Enhanced Raman Spectroscopy with Monolithic Nanoporous Gold Disk Substrates. *Nanoscale* **2013**, *5* (10), 4105.

(91) Santos, G. M.; Zhao, F.; Zeng, J.; Li, M.; Shih, W.-C. Label-Free, Zeptomole Cancer Biomarker Detection by Surface-Enhanced Fluorescence on Nanoporous Gold Disk Plasmonic Nanoparticles. *J. Biophotonics* **2015**, *8* (10), 855–863.

(92) Shih, W. C.; Santos, G. M.; Zhao, F.; Zenasni, O.; Arnob, M. M. P. Simultaneous Chemical and Refractive Index Sensing in the 1-2.5 µm Near-Infrared Wavelength Range on Nanoporous Gold Disks. *Nano Lett.* **2016**, *16* (7), 4641–4647.

(93) Arnob, M. M. P.; Zhao, F.; Zeng, J.; Santos, G. M.; Li, M.; Shih, W. C. Laser Rapid Thermal Annealing Enables Tunable Plasmonics in Nanoporous Gold Nanoparticles. *Nanoscale* **2014**, *6* (21), 12470–12475.

(94) Santos, G. M.; Zhao, F.; Zeng, J.; Shih, W.-C. Characterization of Nanoporous Gold Disks for Photothermal Light Harvesting and Light-Gated Molecular Release. *Nanoscale* **2014**, *6* (11), 5718–5724.

(95) Li, J.; Zhao, F.; Shih, W.-C. Direct-Write Patterning of Nanoporous Gold Microstructures by *in Situ* Laser-Assisted Dealloying. *Opt. Express* **2016**, *24* (20), 23610.

(96) Li, J.; Zhao, F.; Shih, W. C. Photothermal Generation of Programmable Microbubble Array on Nanoporous Gold Disks. *Int. Conf. Opt. MEMS Nanophotonics* **2018**, *2018-July* (13), 18964–18969.

(97) Santos, G. M.; Ferrara, F. I. de S.; Zhao, F.; Rodrigues, D. F.; Shih, W.-C. Photothermal Inactivation of Heat-Resistant Bacteria on Nanoporous Gold Disk Arrays. *Opt. Mater. Express* **2016**, *6* (4), 1217.

(98) Zeng, J.; Zhao, F.; Qi, J.; Li, Y.; Li, C.-H.; Yao, Y.; Lee, T. R.; Shih, W.-C. Internal and External Morphology-Dependent Plasmonic Resonance in Monolithic Nanoporous Gold Nanoparticles. *RSC Adv.* **2014**, *4* (69), 36682–36688.

(99) Zeng, J.; Zhao, F.; Li, M.; Li, C.-H.; Lee, T. R.; Shih, W.-C. Morphological Control and Plasmonic Tuning of Nanoporous Gold Disks by Surface Modifications. *J. Mater. Chem. C* **2015**, *3* (2), 247–252.

(100) Zhao, F.; Zeng, J.; Santos, G. M.; Shih, W.-C. *In Situ* Patterning of Hierarchical Nanoporous Gold Structures by In-Plane Dealloying. *Mater. Sci. Eng. B* **2015**, *194*, 34–40.

(101) Qi, J.; Zeng, J.; Zhao, F.; Lin, S. H.; Raja, B.; Strych, U.; Willson, R. C.; Shih, W.-C.

Label-Free, *in Situ* SERS Monitoring of Individual DNA Hybridization in Microfluidics. *Nanoscale* **2014**, *6* (15), 8521–8526.

(102) Li, M.; Du, Y.; Zhao, F.; Zeng, J.; Mohan, C.; Shih, W.-C. Reagent- and Separation-Free Measurements of Urine Creatinine Concentration Using Stamping Surface Enhanced Raman Scattering (S-SERS). *Biomed. Opt. Express* **2015**, *6* (3), 849.

(103) Li, M.; Zhao, F.; Zeng, J.; Qi, J.; Lu, J.; Shih, W.-C. Microfluidic Surface-Enhanced Raman Scattering Sensor with Monolithically Integrated Nanoporous Gold Disk Arrays for Rapid and Label-Free Biomolecular Detection. *J. Biomed. Opt.* **2014**, *19* (11), 111611.

(104) Qiu, S.; Zhao, F.; Zenasni, O.; Li, J.; Shih, W.-C. Catalytic Assembly of DNA Nanostructures on a Nanoporous Gold Array as 3D Architectures for Label-Free Telomerase Activity Sensing. *Nanoscale Horizons* **2017**, *2* (4), 217–224.

(105) Zhao, F.; Arnob, M. M. P.; Zenasni, O.; Li, J.; Shih, W.-C. Far-Field Plasmonic Coupling in 2-Dimensional Polycrystalline Plasmonic Arrays Enables Wide Tunability with Low-Cost Nanofabrication. *Nanoscale Horizons* **2017**, *2* (5), 267–276.

(106) Chi, H.; Wang, C.; Wang, Z.; Zhu, H.; Mesias, V. S. D.; Dai, X.; Chen, Q.; Liu, W.; Huang, J. Highly Reusable Nanoporous Silver Sheet for Sensitive SERS Detection of Pesticides. *Analyst* **2020**, *145* (15), 5158–5165.

(107) Lee, M. J.; Yang, W. G.; Kim, J. H.; Hwang, K.; Chae, W. S. Silver-Coated Nanoporous Gold Skeletons for Fluorescence Amplification. *Microporous Mesoporous Mater.* **2017**, *237*, 60–64.

(108) Yang, L.; Dou, Y.; Zhou, Z.; Zhang, D.; Wang, S. A Versatile Porous Silver-Coordinated Material for the Heterogeneous Catalysis of Chemical Conversion with Propargylic Alcohols and CO2. *Nanomaterials* **2019**, *9* (11).

(109) Cialone, M.; Celegato, F.; Scaglione, F.; Barrera, G.; Raj, D.; Coïsson, M.; Tiberto, P.; Rizzi, P. Nanoporous FePd Alloy as Multifunctional Ferromagnetic SERS-Active Substrate. *Appl. Surf. Sci.* **2020**, 148759.

(110) Ji, Y.; Xing, Y.; Zhou, F.; Li, X.; Chen, Y.; Shao, L. H. The Mechanical Characteristics of Monolithic Nanoporous Copper and Its Composites. *Adv. Eng. Mater.* **2018**, *20* (10), 1–10.

(111) Diao, F.; Xiao, X.; Luo, B.; Sun, H.; Ding, F.; Ci, L.; Si, P. Two-Step Fabrication of Nanoporous Copper Films with Tunable Morphology for SERS Application. *Appl. Surf. Sci.* **2018**, *427*, 1271–1279.

(112) Zhang, C.; Yue, H.; Wang, H.; Ding, G.; Xiaolin, Z. Nanoporous Copper Films with High Surface Area Formed by Chemical Dealloying from Electroplated CuZn Alloy. *Micro Nanosyst.* **2016**, *8* (1), 13–18.

(113) Wang, N.; Pan, Y.; Wu, S.; Zhang, E.; Dai, W. Fabrication of Nanoporous Copper with Tunable Ligaments and Promising Sonocatalytic Performance by Dealloying Cu-Y Metallic Glasses. *RSC Adv.* **2017**, *7* (68), 43255–43265.

(114) Liu, X.; Li, K.; Chen, M. Substrate for Low Temperature Bonding. **2014**, 223–226.

(115) Hoang, T. T. H.; Verma, S.; Ma, S.; Fister, T. T.; Timoshenko, J.; Frenkel, A. I.; Kenis, P. J. A.; Gewirth, A. A. Nanoporous Copper-Silver Alloys by Additive-Controlled Electrodeposition for the Selective Electroreduction of CO2 to Ethylene and Ethanol. *J. Am. Chem. Soc.* **2018**, *140* (17), 5791–5797.

(116) Liu, W.; Cheng, P.; Yan, J.; Li, N.; Shi, S.; Zhang, S. Temperature-Induced Surface Reconstruction and Interface Structure Evolution on Ligament of Nanoporous Copper. *Sci. Rep.* **2018**, *8* (1), 1–9.

(117) Xu, H.; Pang, S.; Jin, Y.; Zhang, T. General Synthesis of Sponge-Like Ultrafine Nanoporous Metals by Dealloying in Citric Acid. *Nano Res.* **2016**, *9* (8), 2467–2477.

(118) Chan, G. H.; Zhao, J.; Hicks, E. M.; Schatz, G. C.; Van Duyne, R. P. Plasmonic Properties of Copper Nanoparticles Fabricated by Nanosphere Lithography. *Nano Lett.*

**2007**, *7* (7), 1947–1952.

(119) Schade, M.; Fuhrmann, B.; Bohley, C.; Schlenker, S.; Sardana, N.; Schilling, J.; Leipner, H. S. Regular Arrays of Al Nanoparticles for Plasmonic Applications. *J. Appl. Phys.* **2014**, *115* (8), 084309.

(120) Martin, J.; Plain, J. Fabrication of Aluminium Nanostructures for Plasmonics. *J. Phys. D. Appl. Phys.* **2015**, *48* (18), 184002.

(121) Zhu, X.; Imran Hossain, G. M.; George, M.; Farhang, A.; Cicek, A.; Yanik, A. A. Beyond Noble Metals: High Q-Factor Aluminum Nanoplasmonics. *ACS Photonics* **2020**, *7* (2), 416–424.

(122) Rakić, A. D. Algorithm for the Determination of Intrinsic Optical Constants of Metal Films: Application to Aluminum. *Appl. Opt.* **1995**, *34* (22), 4755.

(123) Bisio, F.; Zaccaria, R. P.; Moroni, R.; Maidecchi, G.; Alabastri, A.; Gonella, G.; Giglia, A.; Andolfi, L.; Nannarone, S.; Mattera, L.; Canepa, M. Pushing the High-Energy Limit of Plasmonics. *ACS Nano* **2014**, *8* (9), 9239–9247.

(124) Appusamy, K.; Jiao, X.; Blair, S.; Nahata, A.; Guruswamy, S. Mg Thin Films with Al Seed Layers for UV Plasmonics. *J. Phys. D. Appl. Phys.* **2015**, *48* (18), 184009.

(125) Sterl, F.; Strohfeldt, N.; Walter, R.; Griessen, R.; Tittl, A.; Giessen, H. Magnesium as Novel Material for Active Plasmonics in the Visible Wavelength Range. *Nano Lett.* **2015**, *15* (12), 7949–7955.

(126) Duan, X.; Liu, N. Magnesium for Dynamic Nanoplasmonics. *Acc. Chem. Res.* **2019**, *52* (7), 1979–1989.

(127) Koya, A. N.; Ji, B.; Hao, Z.; Lin, J. Controlling Optical Field Enhancement of a Nanoring Dimer for Plasmon-Based Applications. *J. Opt.* **2016**, *18* (5), 055007.

(128) Maccaferri, N.; Barbillon, G.; Koya, A. N.; Lu, G.; Acuna, G. P.; Garoli, D. Recent Advances in Plasmonic Nanocavities for Single-Molecule Spectroscopy. *Nanoscale Adv.* **2021**, *3* (3), 633–642.

(129) García-Vidal, F. J.; Pendry, J. B. Collective Theory for Surface Enhanced Raman Scattering. *Phys. Rev. Lett.* **1996**, *77* (6), 1163–1166.

(130) Le Ru, E. C.; Blackie, E.; Meyer, M.; Etchegoint, P. G. Surface Enhanced Raman Scattering Enhancement Factors: A Comprehensive Study. *J. Phys. Chem. C* **2007**, *111* (37), 13794–13803.

(131) Purwidyantri, A.; Hsu, C. H.; Yang, C. M.; Prabowo, B. A.; Tian, Y. C.; Lai, C. S. Plasmonic Nanomaterial Structuring for SERS Enhancement. *RSC Adv.* **2019**, *9* (9), 4982–4992.

(132) Okeil, S.; Schneider, J. J. Controlling Surface Morphology and Sensitivity of Granular and Porous Silver Films for Surface-Enhanced Raman Scattering, SERS. *Beilstein J. Nanotechnol.* **2018**, *9* (1), 2813–2831.

(133) Koleva, M. E.; Nedyalkov, N. N.; Atanasov, P. A.; Gerlach, J. W.; Hirsch, D.; Prager, A.; Rauschenbach, B.; Fukata, N.; Jevasuwan, W. Porous Plasmonic Nanocomposites for SERS Substrates Fabricated by Two-Step Laser Method. *J. Alloys Compd.* **2016**, *665*, 282–287.

(134) Ding, Y.; Chen, M. Nanoporous Metals for Catalytic and Optical Applications. *MRS Bull.* **2009**, *34* (8), 569–576.

(135) Li, R.; Liu, X. J.; Wang, H.; Wu, Y.; Chu, X. M.; Lu, Z. P. Nanoporous Silver with Tunable Pore Characteristics and Superior Surface Enhanced Raman Scattering. *Corros. Sci.* **2014**, *84*, 159–164.

(136) Bowden, N.; Brittain, S.; Evans, A. G.; Hutchinson, J. W.; Whitesides, G. M. Spontaneous Formation of Ordered Structures in Thin Films of Metals Supported on an Elastomeric Polymer. *Nature* **1998**, *393* (6681), 146–149.

(137) Lacour, S. P.; Wagner, S.; Huang, Z.; Suo, Z. Stretchable Gold Conductors on Elastomeric Substrates. *Appl. Phys. Lett.* **2003**, *82* (15), 2404–2406.

(138) Liang, X.; Gao, M.; Xie, H.; Xu, Q.; Wu, Y.; Hu, J.; Lu, A.; Zhang, L. Controllable Wrinkling Patterns on Chitosan Microspheres Generated from Self-Assembling Metal Nanoparticles. *ACS Appl. Mater. Interfaces* **2019**, *11* (25), 22824–22833.

(139) Schedl, A. E.; Neuber, C.; Fery, A.; Schmidt, H. W. Controlled Wrinkling of Gradient Metal Films. *Langmuir* **2018**, *34* (47), 14249–14253.

(140) Liu, H.; Zhang, L.; Lang, X.; Yamaguchi, Y.; Iwasaki, H.; Inouye, Y.; Xue, Q.; Chen, M. Single Molecule Detection from a Large-Scale SERS-Active Au 79Ag 21 Substrate. *Sci. Rep.* **2011**, *1*, 1–5.

(141) Ardini M., Huang J. A., Sanchez-Sanchez C., Ponzellini P., Maccaferri N., Jacassi A., Cattarin S., Calandrini E., G. D. Nanoporous Gold Decorated with Silver Nanoparticles as Large Area Efficient SERS Substrate. In *SPIE Conf. Proceeding*; Tanaka, T., Tsai, D. P., Eds.; SPIE, 2017; p 14; San Diego, California, United States.

(142) Jiao, Y.; Ryckman, J. D.; Koktysh, D. S.; Weiss, S. M. Controlling Surface Enhanced Raman Scattering Using Grating-Type Patterned Nanoporous Gold Substrates. *Opt. Mater. Express* **2013**, *3* (8), 1137.

(143) Huang, J.; He, Z.; Liu, Y.; Liu, L.; He, X.; Wang, T.; Yi, Y.; Xie, C.; Du, K. Large Surface-Enhanced Raman Scattering from Nanoporous Gold Film over Nanosphere. *Appl. Surf. Sci.* **2019**, *478* (August 2018), 793–801.

(144) Garoli, D.; Yamazaki, H.; Maccaferri, N.; Wanunu, M. Plasmonic Nanopores for Single-Molecule Detection and Manipulation: Toward Sequencing Applications. *Nano Lett.* **2019**, *19* (11), 7553–7562.

(145) *Far- and Deep-Ultraviolet Spectroscopy*; Ozaki, Y., Kawata, S., Eds.; Springer Japan: Tokyo, 2015.

(146) *Principles of Fluorescence Spectroscopy*; Lakowicz, J. R., Ed.; Springer US: Boston, MA, 2006.

(147) Ahmed, Z.; Agio, M.; Ekinci, Y.; Löffler, J. F.; Jha, S. K. Deep-UV Surface-Enhanced Resonance Raman Scattering of Adenine on Aluminum Nanoparticle Arrays. *J. Am. Chem. Soc.* **2012**, *134* (4), 1966–1969.

(148) Sanz, J. M.; Ortiz, D.; Alcaraz De La Osa, R.; Saiz, J. M.; González, F.; Brown, A. S.; Losurdo, M.; Everitt, H. O.; Moreno, F. UV Plasmonic Behavior of Various Metal Nanoparticles in the Near- and Far-Field Regimes: Geometry and Substrate Effects. *J. Phys. Chem. C* **2013**, *117* (38), 19606–19615.

(149) Ding, T.; Sigle, D. O.; Herrmann, L. O.; Wolverson, D.; Baumberg, J. J. Nanoimprint Lithography of Al Nanovoids for Deep-UV SERS. *ACS Appl. Mater. Interfaces* **2014**, *6* (20), 17358–17363.

(150) Dong, Z.; Wang, T.; Chi, X.; Ho, J.; Tserkezis, C.; Yap, S. L. K.; Rusydi, A.; Tjiptoharsono, F.; Thian, D.; Mortensen, N. A.; Yang, J. K. W. Ultraviolet Interband Plasmonics with Si Nanostructures. *Nano Lett.* **2019**, *19* (11), 8040–8048.

(151) Dörfer, T.; Schmitt, M.; Popp, J. Deep-UV Surface-Enhanced Raman Scattering. *J. Raman Spectrosc.* **2007**, *38* (11), 1379–1382.

(152) Mattiucci, N.; D'Aguanno, G.; Everitt, H. O.; Foreman, J. V; Callahan, J. M.; Buncick, M. C.; Bloemer, M. J. Ultraviolet Surface-Enhanced Raman Scattering at the Plasmonic Band Edge of a Metallic Grating. *Opt. Express* **2012**, *20* (2), 1868–1877.

(153) Sigle, D. O.; Perkins, E.; Baumberg, J. J.; Mahajan, S. Reproducible Deep-UV SERRS on Aluminum Nanovoids. *J. Phys. Chem. Lett.* **2013**, *4* (9), 1449–1452.

(154) Gutierrez, Y.; Ortiz, D.; Sanz, J. M.; Saiz, J. M.; Gonzalez, F.; Everitt, H. O.; Moreno, F. How an Oxide Shell Affects the Ultraviolet Plasmonic Behavior of Ga, Mg, and Al Nanostructures. *Opt. Express* **2016**, *24* (18), 20621.

(155) Yu, P.; Yan, M.; Schaffer, G. B.; Qian, M. Fabrication of Porous Aluminum with Controllable Open-Pore Fraction. *Metall. Mater. Trans. A Phys. Metall. Mater. Sci.* **2011**, *42* (7), 2040–2047.

(156) Hakamada, M.; Kuromura, T.; Chen, Y.; Kusuda, H.; Mabuchi, M. Sound Absorption Characteristics of Porous Aluminum Fabricated by Spacer Method. *J. Appl. Phys.* **2006**, *100* (11).

(157) Vargas-Martínez, J.; Estela-García, J.; Suárez, O.; Vega, C. Fabrication of a Porous Metal *via* Selective Phase Dissolution in Al-Cu Alloys. *Metals (Basel).* **2018**, *8* (6), 378.

(158) Liu, J.; Ni, Z.; Nandi, P.; Mirsaidov, U.; Huang, Z. Chirality Transfer in Galvanic Replacement Reactions. *Nano Lett.* **2019**, *19* (10), 7427–7433.

(159) Garoli, D.; Calandrini, E.; Bozzola, A.; Ortolani, M.; Cattarin, S.; Barison, S.; Toma, A.; De Angelis, F. Boosting Infrared Energy Transfer in 3D Nanoporous Gold Antennas. *Nanoscale* **2017**, *9* (2), 915–922.

(160) Sreekanth, K. V.; Alapan, Y.; Elkabbash, M.; Ilker, E.; Hinczewski, M.; Gurkan, U. A.; De Luca, A.; Strangi, G. Extreme Sensitivity Biosensing Platform Based on Hyperbolic Metamaterials. *Nat. Mater.* **2016**, *15* (6), 621–627.

(161) Calandrini, E.; Giovannini, G.; Garoli, D. 3D Nanoporous Antennas as a Platform for High Sensitivity IR Plasmonic Sensing. *Opt. Express* **2019**, *27* (18), 25912.

(162) Osawa, M. Surface-Enhanced Infrared Absorption. In *Near-Field Optics and Surface Plasmon Polaritons*; Springer Berlin Heidelberg: Berlin, Heidelberg, 2006; pp 163–187.

(163) Anker, J. N.; Hall, W. P.; Lyandres, O.; Shah, N. C.; Zhao, J.; Van Duyne, R. P. Biosensing with Plasmonic Nanosensors. *Nat. Mater.* **2008**, *7* (6), 442–453.

(164) Garoli, D.; Ruffato, G.; Cattarin, S.; Barison, S.; Perino, M.; Ongarello, T.; Romanato, F. Nanoporous Gold—Application to Extraordinary Optical Transmission of Light. *J. Vac. Sci. Technol. B, Nanotechnol. Microelectron. Mater. Process. Meas. Phenom.* **2012**, *31* (1), 012601.

(165) Sherry, L. J.; Jin, R.; Mirkin, C. A.; Schatz, G. C.; Van Duyne, R. P. Localized Surface Plasmon Resonance Spectroscopy of Single Silver Triangular Nanoprisms. *Nano Lett.* **2006**, *6* (9), 2060–2065.

(166) Brown, L. V.; Zhao, K.; King, N.; Sobhani, H.; Nordlander, P.; Halas, N. J. Surface-Enhanced Infrared Absorption Using Individual Cross Antennas Tailored to Chemical Moieties. *J. Am. Chem. Soc.* **2013**, *135* (9), 3688–3695.

(167) Darvill, D.; Centeno, A.; Xie, F. Plasmonic Fluorescence Enhancement by Metal Nanostructures: Shaping the Future of Bionanotechnology. *Phys. Chem. Chem. Phys.* **2013**, *15* (38), 15709.

(168) Li, J.-F.; Li, C.-Y.; Aroca, R. F. Plasmon-Enhanced Fluorescence Spectroscopy. *Chem. Soc. Rev.* **2017**, *46* (13), 3962–3979.

(169) Zhai, Y.-Y.; Liu, Q.; Cai, W.-P.; Cao, S.-H.; Zhang, L.-X.; Li, Y.-Q. Metallic Nanofilm Enhanced Fluorescence Cell Imaging: A Study of Distance-Dependent Intensity and Lifetime by Optical Sectioning Microscopy. *J. Phys. Chem. B* **2020**, *124* (14), 2760–2768.

(170) Guerrero, A. R.; Zhang, Y.; Aroca, R. F. Experimental Confirmation of Local Field Enhancement Determining Far-Field Measurements with Shell-Isolated Silver Nanoparticles. *Small* **2012**, *8* (19), 2964–2967.

(171) Lang, X. Y.; Guan, P. F.; Fujita, T.; Chen, M. W. Tailored Nanoporous Gold for Ultrahigh Fluorescence Enhancement. *Phys. Chem. Chem. Phys.* **2011**, *13* (9), 3795–3799.

(172) Fu, Y.; Zhang, J.; Nowaczyk, K.; Lakowicz, J. R. Enhanced Single Molecule Fluorescence and Reduced Observation Volumes on Nanoporous Gold (NPG) Films. *Chem. Commun.* **2013**, *49* (92), 10874–10876.

(173) Ahmed, S. R.; Hossain, M. A.; Park, J. Y.; Kim, S. H.; Lee, D.; Suzuki, T.; Lee, J.; Park, E. Y. Metal Enhanced Fluorescence on Nanoporous Gold Leaf-Based Assay

Platform for Virus Detection. *Biosens. Bioelectron.* **2014**, *58*, 33–39.

(174) Lang, X. Y.; Guan, P. F.; Fujita, T.; Chen, M. W. Tailored Nanoporous Gold for Ultrahigh Fluorescence Enhancement. *Phys. Chem. Chem. Phys.* **2011**, *13* (9), 3795–3799.

(175) Santos, G. M.; Zhao, F.; Zeng, J.; Li, M.; Shih, W. C. Label-Free, Zeptomole Cancer Biomarker Detection by Surface-Enhanced Fluorescence on Nanoporous Gold Disk Plasmonic Nanoparticles. *J. Biophotonics* **2015**, *8* (10), 855–863.

(176) Yuan, H.; Lu, Y.; Wang, Z.; Ren, Z.; Wang, Y.; Zhang, S.; Zhang, X.; Chen, J. Single Nanoporous Gold Nanowire as a Tunable One-Dimensional Platform for Plasmon-Enhanced Fluorescence. *Chem. Commun.* **2015**, *52* (9), 1808–1811.

(177) Anger, P.; Bharadwaj, P.; Novotny, L. Enhancement and Quenching of Single-Molecule Fluorescence. *Phys. Rev. Lett.* **2006**, *96* (11), 113002.

(178) Bauch, M.; Toma, K.; Toma, M.; Zhang, Q.; Dostalek, J. Plasmon-Enhanced Fluorescence Biosensors: A Review. *Plasmonics* **2014**, *9* (4), 781–799.

(179) Joyce, C.; Fothergill, S. M.; Xie, F. Recent Advances in Gold-Based Metal Enhanced Fluorescence Platforms for Diagnosis and Imaging in the near-Infrared. *Mater. Today Adv.* **2020**, *7*, 100073.

(180) Yuan, H.; Lu, Y.; Wang, Z.; Ren, Z.; Wang, Y.; Zhang, S.; Zhang, X.; Chen, J. Single Nanoporous Gold Nanowire as a Tunable One-Dimensional Platform for Plasmon-Enhanced Fluorescence. *Chem. Commun.* **2016**, *52* (9), 1808–1811.

(181) van der Zalm, J.; Chen, S.; Huang, W.; Chen, A. Review—Recent Advances in the Development of Nanoporous Au for Sensing Applications. *J. Electrochem. Soc.* **2020**, *167* (3), 037532.

(182) Fu, B.; Flynn, J. D.; Isaacoff, B. P.; Rowland, D. J.; Biteen, J. S. Super-Resolving the Distance-Dependent Plasmon-Enhanced Fluorescence of Single Dye and Fluorescent Protein Molecules. *J. Phys. Chem. C* **2015**, *119* (33), 19350–19358.

(183) Knight, M. W.; Liu, L.; Wang, Y.; Brown, L.; Mukherjee, S.; King, N. S.; Everitt, H. O.; Nordlander, P.; Halas, N. J. Aluminum Plasmonic Nanoantennas. *Nano Lett.* **2012**, *12* (11), 6000–6004.

(184) Ding, T.; Sigle, D. O.; Herrmann, L. O.; Wolverson, D.; Baumberg, J. J. Nanoimprint Lithography of Al Nanovoids for Deep-UV SERS. *ACS Appl. Mater. Interfaces* **2014**, *6* (20), 17358–17363.

(185) Zhao, C.; Zhu, Y.; Su, Y.; Guan, Z.; Chen, A.; Ji, X.; Gui, X.; Xiang, R.; Tang, Z. Tailoring Plasmon Resonances in Aluminium Nanoparticle Arrays Fabricated Using Anodic Aluminium Oxide. *Adv. Opt. Mater.* **2015**, *3* (2), 248–256.

(186) McClain, M. J.; Schlather, A. E.; Ringe, E.; King, N. S.; Liu, L.; Manjavacas, A.; Knight, M. W.; Kumar, I.; Whitmire, K. H.; Everitt, H. O.; Nordlander, P.; Halas, N. J. Aluminum Nanocrystals. *Nano Lett.* **2015**, *15* (4), 2751–2755.

(187) Zhu, X.; Daggumati, P.; Seker, E.; Yanik, A. A. Plasmofluidic Nanoporous Gold Membranes for Ultrasensitive Raman Spectroscopy. In *Optical Sensors and Sensing Congress (ES, FTS, HISE, Sensors)*; OSA: Washington, D.C., 2019; p SW5D.4.

(188) Kretschmann, E.; Raether, H. Notizen: Radiative Decay of Non Radiative Surface Plasmons Excited by Light. *Zeitschrift für Naturforsch. A* **1968**, *23* (12), 2135–2136.

(189) Fontana, E.; Pantell, R. H. Characterization of Multilayer Rough Surfaces by Use of Surface-Plasmon Spectroscopy. *Phys. Rev. B* **1988**, *37* (7), 3164–3182.

(190) Raether, H. *Surface Plasmons on Smooth and Rough Surfaces and on Gratings*; Springer Tracts in Modern Physics; Springer Berlin Heidelberg: Berlin, Heidelberg, 1988; Vol. 111.

(191) Porto, J. A.; García-Vidal, F. J.; Pendry, J. B. Transmission Resonances on Metallic Gratings with Very Narrow Slits. *Phys. Rev. Lett.* **1999**, *83* (14), 2845–2848.

(192) Ebbesen, T. W.; Lezec, H. J.; Ghaemi, H. F.; Thio, T.; Wolff, P. A. Extraordinary

Optical Transmission through Sub-Wavelength Hole Arrays. *Nature* **1998**, *391* (6668), 667–669.

(193) Bethe, H. A. Theory of Diffraction by Small Holes. *Phys. Rev.* **1944**, *66* (7–8), 163–182.

(194) Lezec, H. J.; Degiron, A.; Devaux, E.; Linke, R. A.; Martin-Moreno, L.; Garcia-Vidal, F. J.; Ebbesen, T. W. Beaming Light from a Subwavelength Aperture. *Science (80-. ).* **2002**, *297* (5582), 820 LP – 822.

(195) Verslegers, L.; Catrysse, P. B.; Yu, Z.; White, J. S.; Barnard, E. S.; Brongersma, M. L.; Fan, S. Planar Lenses Based on Nanoscale Slit Arrays in a Metallic Film. *Nano Lett.* **2009**, *9* (1), 235–238.

(196) Zhu, X.; Cicek, A.; Li, Y.; Yanik, A. A. Plasmofluidic Microlenses for Label-Free Optical Sorting of Exosomes. *Sci. Rep.* **2019**, *9* (1), 8593.

(197) Brolo, A. G.; Arctander, E.; Gordon, R.; Leathem, B.; Kavanagh, K. L. Nanohole-Enhanced Raman Scattering. *Nano Lett.* **2004**, *4* (10), 2015–2018.

(198) Yanik, A. A.; Huang, M.; Artar, A.; Chang, T.-Y.; Altug, H. Integrated Nanoplasmonic-Nanofluidic Biosensors with Targeted Delivery of Analytes. *Appl. Phys. Lett.* **2010**, *96* (2), 21101.

(199) Eftekhari, F.; Escobedo, C.; Ferreira, J.; Duan, X.; Girotto, E. M.; Brolo, A. G.; Gordon, R.; Sinton, D. Nanoholes as Nanochannels: Flow-Through Plasmonic Sensing. *Anal. Chem.* **2009**, *81* (11), 4308–4311.

(200) Yanik, A. A.; Huang, M.; Kamohara, O.; Artar, A.; Geisbert, T. W.; Connor, J. H.; Altug, H. An Optofluidic Nanoplasmonic Biosensor for Direct Detection of Live Viruses from Biological Media. *Nano Lett.* **2010**, *10* (12), 4962–4969.

(201) Brolo, A. G.; Gordon, R.; Leathem, B.; Kavanagh, K. L. Surface Plasmon Sensor Based on the Enhanced Light Transmission through Arrays of Nanoholes in Gold Films. *Langmuir* **2004**, *20* (12), 4813–4815.

(202) Ruffato, G.; Garoli, D.; Cattarin, S.; Barison, S.; Romanato, F. FIB Lithography of Nanoporous Gold Slits for Extraordinary Transmission. *Microelectron. Eng.* **2012**, *98*, 419–423.

(203) Luk'yanchuk, B.; Zheludev, N. I.; Maier, S. A.; Halas, N. J.; Nordlander, P.; Giessen, H.; Chong, C. T. The Fano Resonance in Plasmonic Nanostructures and Metamaterials. *Nat. Mater.* **2010**, *9* (9), 707–715.

(204) Yanik, A. A.; Cetin, A. E.; Huang, M.; Artar, A.; Mousavi, S. H.; Khanikaev, A.; Connor, J. H.; Shvets, G.; Altug, H. Seeing Protein Monolayers with Naked Eye through Plasmonic Fano Resonances. *Proc. Natl. Acad. Sci.* **2011**, *108* (29), 11784 LP – 11789.

(205) Masson, J.-B.; Gallot, G. Coupling between Surface Plasmons in Subwavelength Hole Arrays. *Phys. Rev. B* **2006**, *73* (12), 121401.

(206) Genet, C.; van Exter, M. P.; Woerdman, J. P. Fano-Type Interpretation of Red Shifts and Red Tails in Hole Array Transmission Spectra. *Opt. Commun.* **2003**, *225* (4), 331–336.

(207) Zhu, X.; Cao, N.; Thibeault, B. J.; Pinsky, B.; Yanik, A. A. Mechanisms of Fano-Resonant Biosensing: Mechanical Loading of Plasmonic Oscillators. *Opt. Commun.* **2020**, *469*, 125780.

(208) Zhu, X.; Daggumati, P.; Seker, E.; Yanik, A. A. Plasmofluidic Nanoporous Gold Membranes for Ultrasensitive Raman Spectroscopy. In *Optical Sensors and Sensing Congress (ES, FTS, HISE, Sensors)*; Optical Society of America, 2019; p SW5D.4; San Jose, California United States.

(209) Kang, E. S. H.; Shiran Chaharsoughi, M.; Rossi, S.; Jonsson, M. P. Hybrid Plasmonic Metasurfaces. *J. Appl. Phys.* **2019**, *126* (14), 140901.

(210) Poddubny, A.; Iorsh, I.; Belov, P.; Kivshar, Y. Hyperbolic Metamaterials. *Nat.*

*Photonics* **2013**, *7* (12), 948–957.
(211) Wu, C.; Khanikaev, A. B.; Adato, R.; Arju, N.; Yanik, A. A.; Altug, H.; Shvets, G. Fano-Resonant Asymmetric Metamaterials for Ultrasensitive Spectroscopy and Identification of Molecular Monolayers. *Nat. Mater.* **2012**, *11* (1), 69–75.
(212) Cortés, E.; Besteiro, L. V.; Alabastri, A.; Baldi, A.; Tagliabue, G.; Demetriadou, A.; Narang, P. Challenges in Plasmonic Catalysis. *ACS Nano* **2020**, acsnano.0c08773.
(213) Moskovits, M. Hot Electrons Cross Boundaries. *Science (80-. ).* **2011**, *332* (6030), 676–677.
(214) Mubeen, S.; Lee, J.; Liu, D.; Stucky, G. D.; Moskovits, M. Panchromatic Photoproduction of H2 with Surface Plasmons. *Nano Lett.* **2015**, *15* (3), 2132–2136.
(215) Linic, S.; Christopher, P.; Ingram, D. B. Plasmonic-Metal Nanostructures for Efficient Conversion of Solar to Chemical Energy. *Nat. Mater.* **2011**, *10* (12), 911–921.
(216) Solís, D. M.; Taboada, J. M.; Obelleiro, F.; Liz-Marzán, L. M.; García De Abajo, F. J. Toward Ultimate Nanoplasmonics Modeling. *ACS Nano* **2014**, *8* (8), 7559–7570.
(217) Zhang, Z.; Zhang, C.; Zheng, H.; Xu, H. Plasmon-Driven Catalysis on Molecules and Nanomaterials. *Acc. Chem. Res.* **2019**, *52* (9), 2506–2515.
(218) Kamarudheen, R.; Aalbers, G. J. W.; Hamans, R. F.; Kamp, L. P. J.; Baldi, A. Distinguishing Among All Possible Activation Mechanisms of a Plasmon-Driven Chemical Reaction. *ACS Energy Lett.* **2020**, *5* (8), 2605–2613.
(219) Ueno, K.; Juodkazis, S.; Shibuya, T.; Yokota, Y.; Mizeikis, V.; Sasaki, K.; Misawa, H. Nanoparticle Plasmon-Assisted Two-Photon Polymerization Induced by Incoherent Excitation Source. *J. Am. Chem. Soc.* **2008**, *130* (22), 6928–6929.
(220) Robatjazi, H.; Zhao, H.; Swearer, D. F.; Hogan, N. J.; Zhou, L.; Alabastri, A.; McClain, M. J.; Nordlander, P.; Halas, N. J. Plasmon-Induced Selective Carbon Dioxide Conversion on Earth-Abundant Aluminum-Cuprous Oxide Antenna-Reactor Nanoparticles. *Nat. Commun.* **2017**, *8* (1), 27.
(221) Swearer, D. F.; Zhao, H.; Zhou, L.; Zhang, C.; Robatjazi, H.; Martirez, J. M. P.; Krauter, C. M.; Yazdi, S.; McClain, M. J.; Ringe, E.; Carter, E. A.; Nordlander, P.; Halas, N. J. Heterometallic Antenna−reactor Complexes for Photocatalysis. *Proc. Natl. Acad. Sci.* **2016**, *113* (32), 8916–8920.
(222) Kim, M.; Lin, M.; Son, J.; Xu, H.; Nam, J. M. Hot-Electron-Mediated Photochemical Reactions: Principles, Recent Advances, and Challenges. *Adv. Opt. Mater.* **2017**, *5* (15), 1–21.
(223) Alabastri, A.; Malerba, M.; Calandrini, E.; Manjavacas, A.; De Angelis, F.; Toma, A.; Proietti Zaccaria, R. Controlling the Heat Dissipation in Temperature-Matched Plasmonic Nanostructures. *Nano Lett.* **2017**, *17* (9), 5472–5480.
(224) Baffou, G.; Quidant, R. Thermo-Plasmonics: Using Metallic Nanostructures as Nano-Sources of Heat. *Laser Photon. Rev.* **2013**, *7* (2), 171–187.
(225) Dubi, Y.; Sivan, Y. "Hot" Electrons in Metallic Nanostructures—Non-Thermal Carriers or Heating? *Light Sci. Appl.* **2019**, *8* (1).
(226) de Barros, H. R.; López-Gallego, F.; Liz-Marzán, L. M. Light-Driven Catalytic Regulation of Enzymes at the Interface with Plasmonic Nanomaterials. *Biochemistry* **2020**, acs.biochem.0c00447.
(227) Biener, M. M.; Biener, J.; Wichmann, A.; Wittstock, A.; Baumann, T. F.; Bäumer, M.; Hamza, A. V. ALD Functionalized Nanoporous Gold: Thermal Stability, Mechanical Properties, and Catalytic Activity. *Nano Lett.* **2011**, *11* (8), 3085–3090.
(228) Wang, Z.; Du, J.; Zhang, Y.; Han, J.; Huang, S.; Hirata, A.; Chen, M. Free-Standing Nanoporous Gold for Direct Plasmon Enhanced Electro-Oxidation of Alcohol Molecules. *Nano Energy* **2019**, *56* (November 2018), 286–293.
(229) Pröschel, A.; Chacko, J.; Whitaker, R.; Chen, M. A. U.; Detsi, E. Visible Light Plasmonic Heating-Enhanced Electrochemical Current in Nanoporous Gold Cathodes.

*J. Electrochem. Soc.* **2019**, *166* (4), H146–H150.

(230) Zhang, X.; Zheng, Y.; Liu, X.; Lu, W.; Dai, J.; Lei, D. Y.; MacFarlane, D. R. Hierarchical Porous Plasmonic Metamaterials for Reproducible Ultrasensitive Surface-Enhanced Raman Spectroscopy. *Adv. Mater.* **2015**, *27* (6), 1090–1096.

(231) Wang, D.; Schaaf, P. Plasmonic Nanosponges. *Adv. Phys. X* **2018**, *3* (1), 478–496.

(232) Larin, A. O.; Nominé, A.; Ageev, E. I.; Ghanbaja, J.; Kolotova, L. N.; Starikov, S. V.; Bruyère, S.; Belmonte, T.; Makarov, S. V.; Zuev, D. A. Plasmonic Nanosponges Filled with Silicon for Enhanced White Light Emission. *Nanoscale* **2020**, *12* (2), 1013–1021.

(233) Zhang, Q.; Large, N.; Nordlander, P.; Wang, H. Porous Au Nanoparticles with Tunable Plasmon Resonances and Intense Field Enhancements for Single-Particle SERS. *J. Phys. Chem. Lett.* **2014**, *5* (2), 370–374.

(234) Kang, T. Y.; Park, K.; Kwon, S. H.; Chae, W. S. Surface-Engineered Nanoporous Gold Nanoparticles for Light-Triggered Drug Release. *Opt. Mater. (Amst).* **2020**, *106* (May), 109985.

(235) Luc, W.; Jiao, F. Nanoporous Metals as Electrocatalysts: State-of-the-Art, Opportunities, and Challenges. *ACS Catal.* **2017**, *7* (9), 5856–5861.

(236) Lang, X.; Hirata, A.; Fujita, T.; Chen, M. Nanoporous Metal/Oxide Hybrid Electrodes for Electrochemical Supercapacitors. *Nat. Nanotechnol.* **2011**, *6* (4), 232–236.

(237) Zhang, J.; Li, C. M. Nanoporous Metals: Fabrication Strategies and Advanced Electrochemical Applications in Catalysis, Sensing and Energy Systems. *Chem. Soc. Rev.* **2012**, *41* (21), 7016–7031.